\documentclass[journal,compsoc]{IEEEtran}
\usepackage[T1]{fontenc}% optional T1 font encoding
\ifCLASSINFOpdf
\else
\fi
\usepackage{amsmath,amssymb,amsfonts}
\usepackage{algorithmic}
\usepackage{graphicx}
\usepackage{textcomp}
\usepackage{color}
\usepackage{caption}  % caption of the table
\usepackage{cite}
\usepackage{pifont}
\usepackage{amsthm}
\usepackage{hyperref}
\usepackage{mathrsfs}
\usepackage{booktabs}
\usepackage{floatpag}
\floatpagestyle{empty}
\usepackage{hyperref}
\usepackage{url}  %Required
\usepackage{graphicx}  %Required
\usepackage[ruled,vlined,linesnumbered]{algorithm2e}
\usepackage{setspace}
\usepackage{multirow}
\usepackage{threeparttable} %表格注释
\usepackage{wrapfig}

\newtheorem{defn}{Definition}

\newtheorem{thm}{Theorem}

\newcommand{\bp}[1]{{\mathbb{P}}\left[{#1}\right]}

\newcommand{\tabincell}[2]{\begin{tabular}{@{}#1@{}}#2\end{tabular}}

\def\BibTeX{{\rm B\kern-.05em{\sc i\kern-.025em b}\kern-.08em T\kern-.1667em\lower.7ex\hbox{E}\kern-.125emX}}

\begin{document}

\title{A Comprehensive Survey on Local Differential Privacy Toward Data Statistics and Analysis}

\author{Teng~Wang,
	Xuefeng~Zhang,
	Jinyu Feng,
	and~Xinyu~Yang
	\IEEEcompsocitemizethanks{\IEEEcompsocthanksitem Teng Wang, Xuefeng Zhang, and Jingyu Feng are with the School of Cyberspace Security, Xi'an University of Posts and Telecommunications. (Corresponding author: Teng Wang. E-mail: wangteng@xupt.edu.cn.) \protect
		\IEEEcompsocthanksitem Xinyu Yang is with the School of Computer Science and Technology, Xi’an Jiaotong University.\protect
}}

\IEEEtitleabstractindextext{%
	\begin{abstract}
		Collecting and analyzing massive data generated from smart devices have become increasingly pervasive in crowdsensing, which are the building blocks for data-driven decision-making. However, extensive statistics and analysis of such data will seriously threaten the privacy of participating users. Local differential privacy (LDP) has been proposed as an excellent and prevalent privacy model with distributed architecture, which can provide strong privacy guarantees for each user while collecting and analyzing data. LDP ensures that each user's data is locally perturbed first in the client-side and then sent to the server-side, thereby protecting data from privacy leaks on both the client-side and server-side. This survey presents a comprehensive and systematic overview of LDP with respect to privacy models, research tasks, enabling mechanisms, and various applications. Specifically, we first provide a theoretical summarization of LDP, including the LDP model, the variants of LDP, and the basic framework of LDP algorithms. Then, we investigate and compare the diverse LDP mechanisms for various data statistics and analysis tasks from the perspectives of frequency estimation, mean estimation, and machine learning. What's more, we also summarize practical LDP-based application scenarios. Finally, we outline several future research directions under LDP.
\end{abstract}

}

\maketitle
\IEEEpeerreviewmaketitle

\section{Introduction}\label{sec-intro}

\IEEEPARstart{W}{ith} the rapid development of wireless communication techniques, Internet-connected devices (e.g., smart devices and IoT appliances) are ever-increasing and generate large amounts of data by crowdsensing \cite{cheng2017mobile}. Undeniably, these big data have brought our rich knowledge and enormous benefits, which deeply facilitates our daily lives, such as traffic flow control, epidemic prediction, and recommendation systems\cite{guo2015mobile,shu2018privacy,lu2019study}. To make better collective decisions and improve service quality, a variety of applications collect users data through crowdsensing to analyze statistical knowledge of the social community \cite{jarrett2018crowdsourcing}. For example, the third-parties learn rating aggregation by gathering preference options \cite{ krzywicki2015collaborative}, present a crowd density map by recording users locations \cite{chen2016private}, and estimate the power usage distributions from meter readings \cite{yao2014efficient,liu2018practical}. Almost all data statistics and analysis tasks fundamentally depend on a basic understanding of the distribution of the data.

However, collecting and analyzing data has incurred serious privacy issues since such data contain various sensitive information of users \cite{fung2010privacy,zhu2017differentially,yang2017survey2}. What's worse, driven by advanced data fusion and analysis techniques, the private data of users are more vulnerable to attack and disclosure in the big data era \cite{soria2016big,yu2016big,sun2020privacy}. For example, the adversaries can infer the daily habits or behavior profiles of family members (e.g., the time of presence/absence in the home, the certain activities such as watching TV, cooking) by analyzing the usage of appliances \cite{hino2013versatile, zhao2014achieving, barbosa2016technique}, and even obtain the identification information, social relationships, religion attitudes \cite{wang2019privacy}.

Therefore, it's an urgent priority to put great attention on preventing personal data from being leaked when collecting data from various devices. At present, the European Union (EU) has published the GDPR \cite{GDPR} that regulates the EU laws of data protection for all individual citizens and contains the provisions and requirements pertaining to the processing of personal data. Besides, the NIST of the U.S. is also developing the privacy frameworks \cite{NIST} currently to better identify, access, manage, and communicate about privacy risks so that individuals can enjoy the benefits of innovative technologies with greater confidence and trust.

From the perspective of privacy-preserving techniques, differential privacy (DP) \cite{dwork2014algorithmic} has been proposed for more than ten years and recognized as a convincing framework for privacy protection, which also refers to global DP (or centralized DP)\footnote{Without loss of generality, DP appears in the rest of this article refers to global DP (i.e., centralized DP).}. With strict mathematical proofs, DP is independent of the background knowledge of adversaries and capable of providing each user with strong privacy guarantees, which has been widely adopted and used in many areas \cite{yang2017survey,abowd2018us}. However, DP can be only used to the assumption of a trusted server. In many online services or crowdsourcing systems, the servers are untrustworthy and always interested in the statistics of users' data. 

Based on the definition of DP, local differential privacy (LDP) \cite{kasiviswanathan2011can} is proposed as a distributed variant of DP, which achieves privacy guarantees for each user locally and is independent of any assumptions on the third-party servers. LDP has been imposed as the cutting-edge of research on privacy protection and risen in prominence not only from theoretical interests, but also subsequently from a practical perspective. For example, many companies have deployed LDP-based algorithms in real systems, such as Apple iOS \cite{apple2017local}, Google Chrome \cite{erlingsson2014rappor}, Windows system \cite{ding2017collecting}.

Due to its powerfulness, LDP has been widely adopted to alleviate the privacy concerns of each user while conducting statistical and analytic tasks, such as frequency and mean value estimation \cite{wang2019collecting}, heavy hitters discovery \cite{qin2016heavy}, $k$-way marginal release \cite{cormode2018marginal}, empirical risk minimization (ERM) \cite{wang2018empirical}, federated learning \cite{wang2020federated}, and deep learning \cite{arachchige2019local}. 

Therefore, a comprehensive survey of LDP is very necessary and urgent for future research in Internet of Things. To the best of our knowledge, only a little literature focuses on reviewing LDP and the most existing surveys only pay attention to a certain field. For example, Wang \textit{et al.} \cite{wang2017protocol} summarized several LDP protocols only for frequency estimation. The tutorials in  \cite{ye2019local,bebensee2019local,li2019mobile} reviewed the LDP models and introduced the current research landscapes under LDP, but the detailed descriptions are rather insufficient. Zhao \textit{et al.} \cite{zhao2019survey} reviewed the existing LDP-based mechanisms only towards the Internet of connected vehicles. The reviews in \cite{yang2020local,xiong2020comprehensive} also provided a survey of statistical query and private learning with LDP. However, the detailed technical points and specific data types when using LDP are still insufficiently summarized. Therefore, it is still necessary and urgent to carry out a comprehensive survey on LDP toward data statistics and analysis to help newcomers understand the complex discipline of this hot research area.

In this survey, we conduct an in-depth overview of LDP with respect to its privacy models, the related research tasks for various data, enabling mechanisms, and wide applications. Our main contributions are summarized as follows.
\begin{enumerate}
	\item We firstly provide a theoretical summarization of LDP from the perspectives of the LDP models, the general framework of LDP algorithms, and the variants of LDP.

	\item We systematically investigate and summarize the enabling LDP mechanisms for various data statistics and analysis tasks. In particular, the existing state-of-the-art LDP mechanisms are thoroughly concluded from the perspectives of frequency estimation, mean value estimation, and machine learning.

	\item We explore the practical applications with LDP to show how LDP is to be implemented in various applications, including in real systems (e.g., Google Chrome, Apple iOS), edge computing, hypothesis testing, social networks, and recommendation systems.
	 
	\item We further distinguish some promising research directions of LDP, which can provide useful guidance for new researchers.
\end{enumerate}

\begin{figure*}[ht]\centering
	\includegraphics[width=1\textwidth]{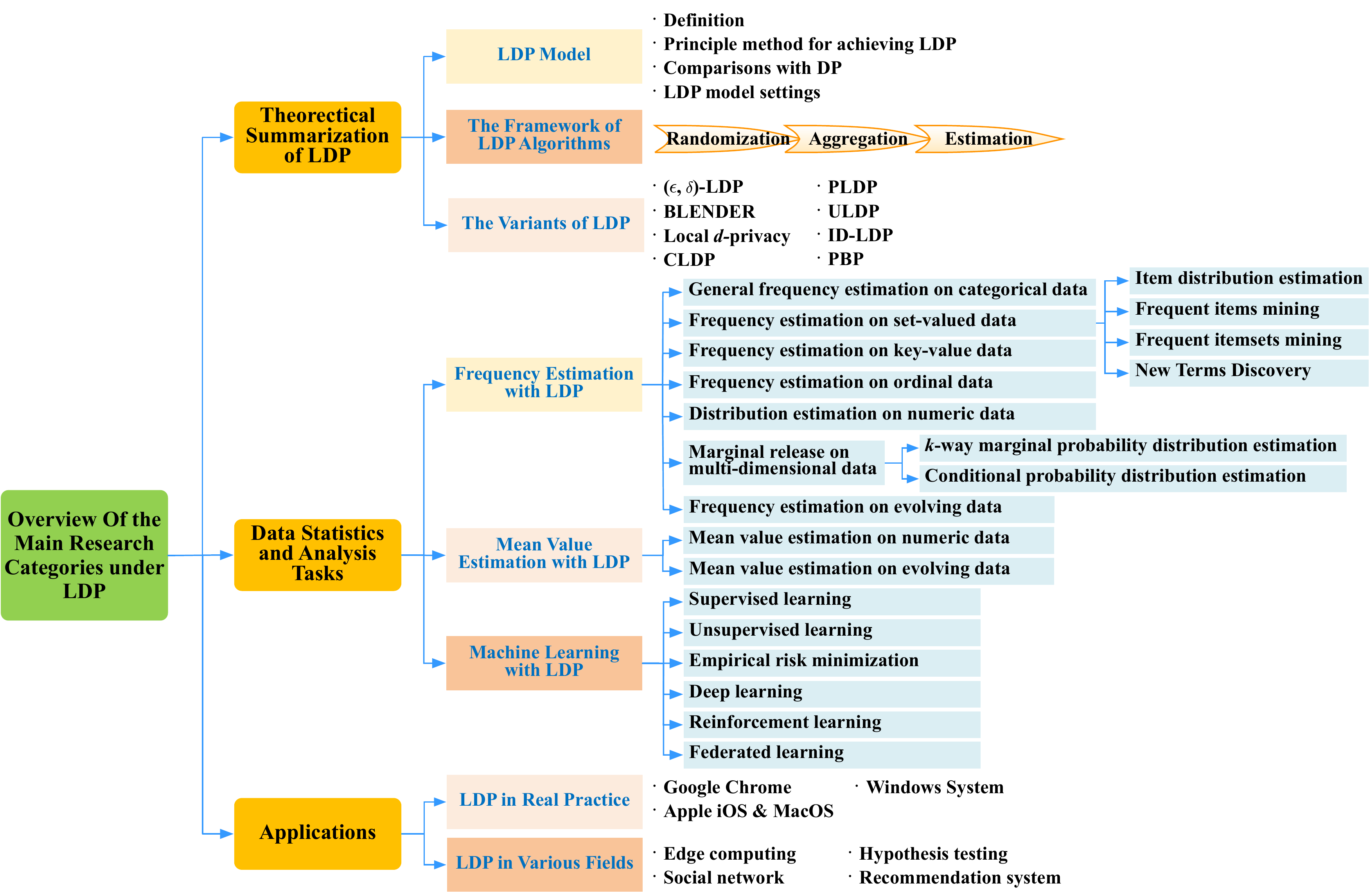}
	\caption{An overview of the main research categories with LDP}\label{fig-task}
	\vspace{-2mm}
\end{figure*}

Fig.~\ref{fig-task} presents the main research categories of LDP and also shows the main structure of this survey. We first provide a theoretical summarization of LDP, which includes the LDP model, the framework of LDP algorithms, and the variants of LDP. Then, from the perspective research tasks, we summarize the existing LDP-based privacy-preserving mechanisms into three categories: \textbf{frequency estimation}, \textbf{mean estimation} and \textbf{machine learning}. We further subdivide each category into several subtasks based on different data types. In addition, we summarize the applications of LDP in real practice and other fields.

The rest paper is organized as follows. Section~\ref{sec-preliminaries} theoretically summarizes the LDP. The diverse LDP mechanisms for frequency estimation, mean estimation and machine learning are introduced thoroughly in Section~\ref{sec-fre}, Section~\ref{sec-mean}, and Section~\ref{sec-ml}, respectively. Section~\ref{sec-application} summarizes the wide application scenarios of LDP and Section~\ref{sec-future} presents some future research directions. Finally, we conclude the paper in Section~\ref{sec-conclusion}.

%\begin{figure}[t]\centering
%	\includegraphics[width=.4\textwidth]{figures/LDP_bg.pdf}
%	\caption{The high-level privacy model under LDP}
%	\label{comp-var}
%\end{figure}

\section{Theoretical Summarization of LDP}\label{sec-preliminaries}

Formally, let $N$ be the number of users, and $U_i(1\leq i \leq N)$ denote the $i$-th user. Let $V^i$ denote the data record of $U_i$, which is sampled from the attribute domain $\mathcal{A}$ that consists of $d$ attributes $A_1,A_2,\cdots,A_d$. For categorical attribute, its discrete domain is denoted as $\mathcal{K}=\{v_1,v_2,\cdots,v_k\}$, where $k$ is the size of the domain and $|\mathcal{K}|=k$. Notations commonly used in this paper are listed in Table~\ref{notations}.  %We will explain other specific notations at where they appear for the first time.

\begin{table}[t]
\caption{The commonly used notations}\label{notations}
\centering
\footnotesize
\setlength{\tabcolsep}{1mm}{
\begin{tabular}{l|l}
\bottomrule[0.8pt]
\textbf{Notation} & \textbf{Explanation} \\
\hline
$V^i$   &Data record of user $U_i$\\
$\mathcal{K}=\{v_1,\cdots,v_k\}$ &Domain of categorical attribute with size $|\mathcal{K}|=k$  \\
$v$~/~$v^*$     &Input value / Perturbed value    \\
$B$           &Vector of the encoded value \\
$N$           &Number of users        \\
$N_v$ / $\bar{N}_v$ / $\widehat{N}_v$  &The true/reported/estimated number of value $v$ \\
$A$      &Attribute  \\
$d$           &Dimension   \\
$\epsilon$/$\delta$  &Privacy budget/Probability of failure \\
$p,q$         &Perturbation probability   \\
%$c_v$ / $\widehat{c}_v$     &The true/estimated count of value $v$            \\
$f_v$ / $\widehat{f}_v$     &The true/estimated frequency of value $v$            \\
$\mathbb{H}$ / $H$  &Hash function universe / Hash function \\
\toprule[1pt]
\end{tabular}}
\end{table}

\subsection{LDP Model}

Local differential privacy is a distributed variant of DP. It allows each user to report her/his value $v$ locally and send the perturbed data to the server aggregator. Therefore, the aggregator will never access to the true data of each user, thus providing a strong protection. Here, user's value $v$ acts as the input value of a perturbation mechanism and the perturbed data acts as the output value.

\subsubsection{Definition}

\begin{defn}[$\epsilon$-Local Differential Privacy ($\epsilon$-LDP) \cite{kasiviswanathan2011can,duchi2013local}] \label{defn-eps-ldp}
A randomized mechanism $\mathcal{M}$ satisfies $\epsilon$-LDP if and only if for any pairs of input values $v$, $v'$ in the domain of $\mathcal{M}$, and for any possible output $y\in\mathcal{Y}$, it holds
\begin{align}\label{eqn-eps-ldp}
\mathbb{P}[\mathcal{M}(v) = y] \leq e^\epsilon \cdot \mathbb{P}[\mathcal{M}(v') = y],
\end{align}
where $\mathbb{P}[\cdot]$ denotes probability and $\epsilon$ is the privacy budget. A smaller $\epsilon$ means stronger privacy protection, and vice versa.
\end{defn}

%Central differential privacy has two important composition properties, that is sequential composition and parallel composition. Although the definitions of neighboring datasets of central DP and LDP are different, the privacy protection paradigms are the same in essential. Therefore, local differential privacy also holds the above two composition properties which are formally defined as follows.

Sequential composition is a key theorem of LDP, which plays important roles in some complex LDP algorithms or some complex scenarios.

\begin{thm}[Sequential Composition]
	\label{compo-sequence}
	Let $\mathcal{M}_i(v)$ be an $\epsilon_i$-LDP algorithm on an input value $v$, and $\mathcal{M}(v)$ is the sequential composition of $\mathcal{M}_1(v), ..., \mathcal{M}_m(v)$. Then $\mathcal{M}(v)$ satisfies $\sum_{i=1}^m \epsilon_i$-LDP.
\end{thm}

%Note that DP has two composition properties, that is sequential composition and parallel composition. Under LDP, parallel composition

%\begin{thm}[Parallel Composition]
%	\label{compo-para}
%	Let $t_1,t_2,\cdots,t_m$ be a set of disjoint subsets of the tuple $t$, where each subset $t_i$ is provided $\epsilon _i$-LDP by algorithm ${\mathcal A}_i$. Then, the algorithms ${\mathcal A}_1,{\mathcal A}_2, \cdots, {\mathcal A}_m$ provide $\underset{i}\max~\epsilon _i$-LDP for tuple $t$.
%\end{thm}

\subsubsection{The Principle Method for Achieving LDP}\label{sec-randomized-response}

\textbf{Randomized response} (RR) \cite{warner1965randomized} is the classical technique for achieving LDP, which can also be used for achieving global DP \cite{wang2016using}. The main idea of RR is to protect user's private information by answering a plausible response to the sensitive query. That is, one user who possesses a private bit $x$ flips it with probability $p$ to give the true answer and with probability $1-p$ to give other answers.

For example, the data collector wants to count the true proportion $f$ of the smoker among $N$ users. Each user is required to answer the question ``Are you a smoker?'' with ``Yes'' or ``No''. To protect privacy, each user flips an unfair coin with the probability $p$ being head and the probability $1-p$ being tail. If the coin comes up to the head, the user will respond the true answer. Otherwise, the user will respond the opposite answer. In this way, the probabilities of answering ``Yes'' and ``No'' can be calculated as
\begin{align}
    &\bp{\text{answer=``Yes''}}=fp+(1-f)(1-p),\label{prob-yes}\\
    &\bp{\text{answer=``No''}}=(1-f)p+f(1-p).\label{prob-no}
\end{align}
Then, we estimate the number of ``Yes'' and ``No'' that are denoted as $N_1$ and $N-N_1$. From Eqs.~(\ref{prob-yes}) and (\ref{prob-no}), we have $N_1/N=fp+(1-f)(1-p)$ and $(N-N_1)/N=(1-f)p+f(1-p)$. Therefore, we can compute the estimated proportion of the smoker is
\begin{align}
	\widehat{f}=\frac{p-1}{2p-1}+\frac{N_1}{(2p-1)N}.
\end{align}

Observe that in the above example the probability of receiving ``Yes'' varies from $p$ to $1-p$ depending on the true information of users. Similarly, the probability of receiving ``No'' also varies from $p$ to $1-p$. Hence, the ratio of probabilities for different answers of one user can be at most $\frac{p}{1-p}$. By letting $\frac{p}{1-p}=e^\epsilon$ and based on Eq.~(\ref{eqn-eps-ldp}), it can be easily verified that the above example satisfies $(\ln\frac{p}{1-p})$-LDP. To ensure  $(\ln\frac{p}{1-p})>0$, we should make sure that $p>\frac{1}{2}$.

Therefore, RR achieves LDP by providing plausible deniability for the responses of users. In this case, users no longer need to trust a centralized curator since they only report plausible data. Based on RR, there are plenty of other mechanisms for achieving LDP under different research tasks, which will be introduced in the following Sections.

\begin{figure*}[t]\centering
	\includegraphics[width=.8\textwidth]{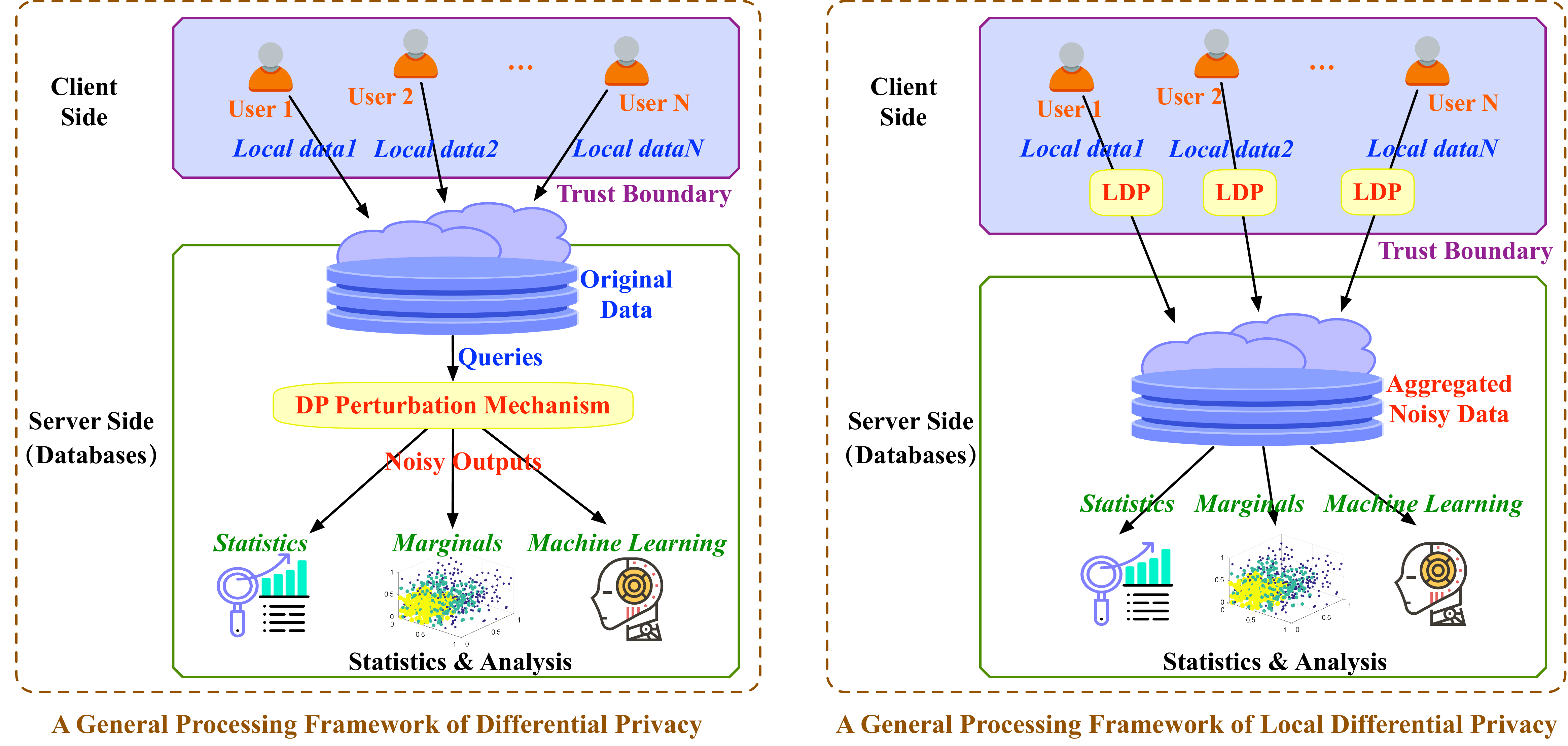}
	\caption{The general processing frameworks of DP and LDP}
	\label{DPvsLDP}
\end{figure*}

\subsubsection{Comparisons with Global Differential Privacy}

We compare LDP with global DP from different perspectives, as shown in Table~\ref{compare-dp-ldp}. At first, the biggest difference between LDP and DP is that LDP is a local privacy model with no assumption on the server while DP is a central privacy model with the assumption of a trusted server. Correspondingly, the general processing frameworks of DP and LDP are different. As shown in the left part of Fig.~\ref{DPvsLDP}, under the DP framework, the data are directly sent to the server and the noises are added to query mechanisms in the server-side. In contrast, under the LDP framework, each user's data are locally perturbed in the client-side before uploading to the server, as shown in the right part of Fig.~\ref{DPvsLDP}.

\begin{table*}[htb]
	\centering
	\footnotesize
	\renewcommand\arraystretch{1.5} % 表格行距
	\caption{Comparisons between LDP and DP}\label{compare-dp-ldp}
	\setlength{\tabcolsep}{0.5mm}{
		\begin{tabular}{c|c|c|c|c|c|c}
			\bottomrule[1pt]
			% 第一行
			\textbf{\tabincell{c}{Notion}} 
			&\textbf{\tabincell{c}{Model}}  
			&\textbf{\tabincell{c}{Server}}
			&\textbf{\tabincell{c}{Neighboring Datasets}}
			&\textbf{Basic Mechanism} 
			&\textbf{Property} 
			&\textbf{Applications} \\ 
			\hline
			% 第二行
			\textbf{DP} \cite{dwork06Calibrating,dwork2014algorithmic}
			&Central  
			&Trusted
			&Two datasets
			&\tabincell{l}{Laplace/Exponential\\Mechanisms \cite{dwork06Calibrating,mcsherry2007mechanism}}
			&\multirow{2}{*}{\tabincell{l}{Sequential\\Composition,\\Post-processing}}  
			&\multirow{2}{*}{\tabincell{l}{Data collection,\\statistics, publishing,\\analysis}}\\ 
			\cline{1-5}
			% 第三行
			\textbf{LDP} \cite{kasiviswanathan2011can,duchi2013local} &Local    
			&\tabincell{l}{No requirement} 
			&Two records
			&\tabincell{l}{Randomized\\Response \cite{warner1965randomized,duchi2013local}} 
			&   
			& \\
			\toprule[1pt]
		\end{tabular}
	}
\end{table*}

The neighboring datasets in LDP are defined as two different records/values of the input domain. While in DP, the neighboring datasets are defined as two datasets that differ only in one record. For example, given a dataset, we can get its neighboring dataset by deleting/modifying one record. The most two common perturbation mechanisms of achieving DP are Laplace mechanism and Exponential mechanism \cite{dwork06Calibrating,mcsherry2007mechanism} that inject random noises based on privacy budget and sensitivity. In contrast, the randomized response technique \cite{warner1965randomized,duchi2013local} is most commonly used to achieve LDP. As shown in Table~\ref{compare-dp-ldp}, LDP holds the same sequential composition and post-processing properties as DP. Both DP and LDP are widely adopted by many applications, such as data collection, publishing, analysis, and so on.

\subsubsection{LDP Model Settings}
This section summarizes the model settings of LDP, which holds two paradigms, that is interactive setting and non-interactive setting \cite{duchi2013local, duchi2013local-long}.

%Recall that in central DP, there are two privacy model settings \cite{yang2017survey}, namely interactive and non-interactive. Since DP assumes a trusted third-party data curator (or, data collector), thus the data curator will collect all the data records denoted as dataset $D$ and then response to the query requests of data analysts. For interactive setting of DP, the data curator returns the noisy query outputs to the data analysts whenever they submit the queries. For non-interactive setting of DP, data curator directly publishes noisy statistical information from which the data analysts execute the queries and obtain the noisy outputs.

\begin{figure}[t]\centering
	\includegraphics[width=.45\textwidth]{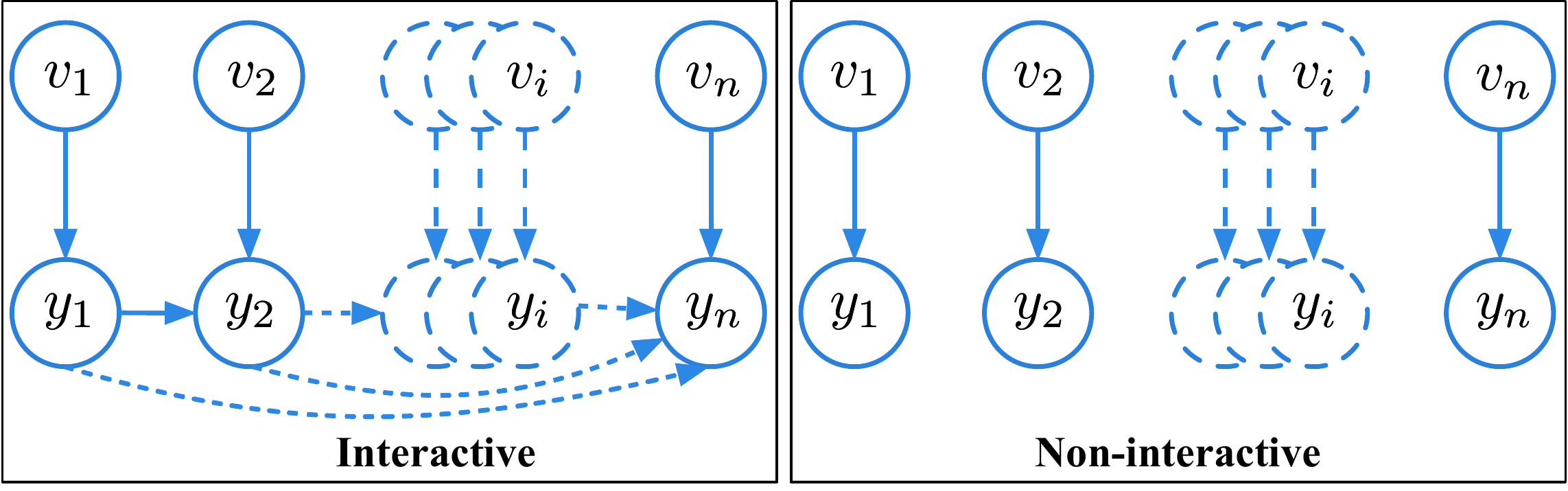}
	\caption{LDP model settings}
	\label{interactive}
	\vspace{-1mm}
\end{figure} 

Since LDP no longer assumes a trusted third-party data curator, the interactive and non-interactive privacy model settings of LDP \cite{duchi2013local-long} are different from that of DP \cite{yang2017survey}. Fig.~\ref{interactive} shows the interactive and non-interactive settings of LDP.

Let $v_1,v_2,\cdots,v_n \in \mathcal{K}$ be the input sequences, and $y_1,y_2,\cdots,y_n \in \mathcal{Y}$ be the corresponding output sequences. As shown in left part of Fig.~\ref{interactive}, in interactive setting, the $i$-th output $y_i$ depends on the $i$-th input $v_i$ and the previous $i-1$ outputs $y_{1:i-1}$, but is independent of the previous $i-1$ inputs $v_{1:i-1}$. Particularly, the dependence and conditional independence correlations can be formally denoted as $\{v_i,y_1,\cdots,y_{i-1}\} \rightarrow y_i \wedge y_i \perp v_j | \{v_i,y_1,\cdots,y_{i-1}\}$ for any $j \neq i$.

In contrast, as shown in the right part of Fig.~\ref{interactive}, the non-interactive setting is much simpler than interactive setting. The $i$-th output $y_i$ only depends on the $i$-th input $v_i$. In formal, the dependence and conditional independence correlations can be denoted as $ v_i \rightarrow y_i \wedge y_i \perp \{v_j,y_j,j\neq i\}|v_i$.

Therefore, the main difference between interactive and non-interactive settings of LDP is whether to consider the correlations between the output results. The work in \cite{joseph2019role,wang2019sparse} further investigated the power of interactivity in LDP.

\subsection{The Framework of LDP Algorithm}\label{sec-framework}

The general privacy-preserving framework with LDP includes three modules: \textsf{Randomization}, \textsf{Aggregation} and \textsf{Estimation}, as shown in Algorithm~\ref{algorithm1}. The randomization is conducted in the client side and both aggregation and estimation happen in the server side.

\begin{algorithm}\small
	\caption{\small{The General Procedure of LDP-based Privacy-preserving Mechanisms}}\label{algorithm1}
	\KwIn{original data of the users, privacy parameter $\epsilon$}
	\KwOut{objective information $f$}
	\tcc{Randomization in the client side}
	\For{each user $U_i$ $(i \in [1,N])$ with value $v$}
	{   Encode input value when necessary\;
		Conduct perturbation with privacy parameter $\epsilon$\;
		Report noisy value $\widehat{v}$ to the server\;
	}
	\tcc{Aggregation in the server side}
	Aggregate all reported values\;
	\tcc{Estimation in the server side}
	Estimate and correct the objective information $f$\;
	\textbf{return} $f$
\end{algorithm}

\subsection{The Variants of LDP}\label{sec-variant}

Since the  introduction of LDP, designing the variant of LDP has been an important research direction to improve the utility of LDP and to make LDP more relevant in targeted IoT scenarios. This section summarizes the current research progresses on LDP variant, as shown in Table~\ref{summary-ldp-variant}.

\subsubsection{$(\epsilon,\delta)$-LDP}

Similar to the case that $(\epsilon, \delta)$-DP \cite{dwork2006our} is a relaxation   of $\epsilon$-DP,   $(\epsilon, \delta)$-LDP (also called \textit{approximate} LDP) is a relaxation of $\epsilon$-LDP (also called \textit{pure} LDP). 
% And $(\epsilon, \delta)$-local differential privacy can achieve better accuracy than $\epsilon$-local differential privacy when applied to deep learning algorithms.

\begin{defn}[$(\epsilon, \delta)$-Local Differential Privacy ($(\epsilon, \delta)$-LDP) \cite{bassily2019linear}] \label{defn-eps-delta-ldp}
	A randomized mechanism $\mathcal{M}$ satisfies $(\epsilon, \delta)$-LDP if and only if for any pairs of input values $v$ and $v'$ in the domain of $\mathcal{M}$, and for any possible output $y\in\mathcal{Y}$, it holds
	\begin{align}\label{eqn-eps-delta-ldp}
	\mathbb{P}[\mathcal{M}(v) = y] \leq e^\epsilon \cdot \mathbb{P}[\mathcal{M}(v') = y] + \delta,
	\end{align}
	where $\delta$ is typically small.
\end{defn}

Loosely speaking, \mbox{$(\epsilon,\delta)$-LDP} means that a mechanism $\mathcal{M}$ achieves \mbox{$\epsilon$-LDP} with probability at least $1-\delta$. By relaxing $\epsilon$-LDP,    $(\epsilon, \delta)$-LDP  is more general since the latter in the special case of $\delta=0$ becomes the former.

\subsubsection{BLENDER}

BLENDER \cite{avent2017blender} is a hybrid model by combining global DP and LDP, which improves data utility with desired privacy guarantees. The BLENDER is achieved by separating the user pool into two groups based on their trust in the data aggregator. One is called \textit{opt-in group} that contains the users who have higher trust in the aggregator. Another is called \textit{clients} that contains the remaining users. Then, the BLENDER can maximize the data utility by balancing the data obtained from participation of opt-in users with that of other users. The privacy definition of BLENDER is the same as $(\epsilon, \delta)$-DP \cite{dwork2006our}.

\subsubsection{Local $\textsf{d}$-privacy}

Geo-indistinguishability \cite{andres2013geo} was initially proposed for location privacy protection under global DP, which is defined based on the geographical distance of data. Geo-indistinguishability  has been quite successful when the statistics are distance-sensitive. In the local settings, Alvim \textit{et al.} \cite{alvim2018metric} also pointed out that the metric-based LDP can provide better utility that standard LDP. Therefore, based on $\textsf{d}$-privacy \cite{chatzikokolakis2013broadening}, Alvim \textit{et al.} \cite{alvim2018metric} proposed local $\textsf{d}$-privacy that is as defined as follows.

\begin{defn}[Local $\textsf{d}$-Privacy] 
	A randomized mechanism $\mathcal{M}$ satisfies local $\textsf{d}$-privacy if and only if for any pairs of input values $v$ and $v'$ in the domain of $\mathcal{M}$, and for any possible output $y\in\mathcal{Y}$, it holds
	\begin{align}
	\mathbb{P}[\mathcal{M}(v) = y] \leq e^{\epsilon\cdot \textsf{d}(v,v')} \cdot \mathbb{P}[\mathcal{M}(v') = y],
	\end{align}
	where $\textsf{d}(\cdot,\cdot)$ is a distance metric.
\end{defn}

Local $\textsf{d}$-privacy can relax the privacy constraint by introducing a distance metric when $\textsf{d}(v,v') >1$, thus improving data utility. In other words, the relaxation of local $\textsf{d}$-privacy is reflected in that the two data becomes more distinguishable as their distance increases. Therefore, local $\textsf{d}$-privacy is quite appropriate for distance-sensitive data, such as location data, energy consumption in smart meters.

\subsubsection{CLDP}

LDP has played an important role in data statistics and analysis. However, the standard LDP will suffer a poorly data utility when the number of users is small. To address this, Gursoy \textit{et al.} \cite{gursoy2019secure} introduced condensed local differential privacy (CLDP) that is also a metric-based privacy notation. Let $\textsf{d}(\cdot,\cdot)$ be a distance metric. Then, CLDP is defined as follows.

\begin{defn}[$\alpha$-CLDP] 
	A randomized mechanism $\mathcal{M}$ satisfies $\alpha$-CLDP if and only if for any pairs of input values $v$ and $v'$ in the domain of $\mathcal{M}$, and for any possible output $y\in\mathcal{Y}$, it holds
	\begin{align}
	\mathbb{P}[\mathcal{M}(v) = y] \leq e^{\alpha\cdot \textsf{d}(v,v')} \cdot \mathbb{P}[\mathcal{M}(v') = y],
	\end{align}
	where $\alpha >0$.
\end{defn}

By definition, in CLDP, $\alpha$ must decrease to compensate as distance $\textsf{d}$ increases. Thus, it holds that $\alpha \ll \epsilon$. In addition, Gursoy \textit{et al.} \cite{gursoy2019secure} also adopted a variant of the Exponential Mechanism (EM) to design several protocols that achieve CLDP with better data utility when there is a small number of users.

\subsubsection{PLDP}
Instead setting a global privacy constraint for all users, personalized local differential privacy (PLDP) \cite{chen2016private,nie2019utility} is proposed to provide granular privacy constraints for each participating user. That is, under PLDP, each user can select the privacy demand (i.e., $\epsilon$) according to his/her own preference. In formal, PLDP is defined as follows.

\begin{defn}[$\epsilon$-PLDP]
	A randomized mechanism $\mathcal{M}$ satisfies $\epsilon_U$-PLDP if and only if for any pairs of input values $v$ and $v'$ in the domain of $\mathcal{M}$ and a user $U$, and for any possible output $y\in\mathcal{Y}$, it holds
	\begin{align}
	\mathbb{P}[\mathcal{M}(v) = y] \leq e^{\epsilon_U} \cdot \mathbb{P}[\mathcal{M}(v') = y],
	\end{align}
	where $\epsilon_U$ is the privacy budget belonging to user $U$.
\end{defn}

To achieve PLDP, Chen \textit{et al.} \cite{chen2016private} proposed personalized count estimation (PCE) protocol and further leveraged a user group clustering algorithm to apply PCE to users with different privacy level. In addition, Nie \textit{et al.} \cite{nie2019utility} proposed the advanced combination strategy to compose multilevel privacy demand with an optimal utility.

\subsubsection{ULDP}

The standard LDP regards all user data equally sensitive, which leads to excessive perturbations. In fact, not all personal data are equally sensitive. For example, answer a questionnaires such as: ``Are you a smoker?'' Obviously, ``Yes'' is a sensitive answer, whereas ``No'' is not sensitive. To improve data utility, Utility-optimized LDP (ULDP) \cite{murakami2019utility} was proposed as a new privacy notation to provide privacy guarantees only for sensitive data. In ULDP, let $\mathcal{K}_S \subseteq \mathcal{K}$ be the sensitive data set, and $\mathcal{K}_N = \mathcal{K}\setminus \mathcal{K}_S$ be the remaining data set. Let $\mathcal{Y}_P \subseteq \mathcal{Y}$ be the protected data set, and $\mathcal{Y}_I = \mathcal{Y} \setminus \mathcal{Y}_P$ be the invertible data set. Then, ULDP is formally defined as follows.

\begin{defn}[$(\mathcal{K}_S, \mathcal{Y}_P, \epsilon)$-ULDP]\label{def-ULDP}
	Given $\mathcal{K}_S \subseteq \mathcal{K}$,  $\mathcal{Y}_P \subseteq \mathcal{Y}$, a randomized mechanism $\mathcal{M}$ provides $(\mathcal{K}_S, \mathcal{Y}_P, \epsilon)$-PLDP if it satisfies the following properties:
	
	(i) For any $y\in \mathcal{Y}_I$, there exists an $v\in \mathcal{X}_N$ such that
	\begin{align}\label{uldp-1}
		\mathbb{P}[\mathcal{M}(v)=y]>0, \mathbb{P}[\mathcal{M}(v')=y]=0 \text{~for~any~} x'\neq x  
	\end{align}
	
	(ii) For any $v,v' \in \mathcal{K}$ and any $y\in \mathcal{Y}_P$,
	\begin{align}\label{uldp-2}
		\mathbb{P}[\mathcal{M}(v) = y] \leq e^\epsilon \cdot \mathbb{P}[\mathcal{M}(v') = y]
	\end{align}
\end{defn}

For an intuitive understanding for Definition \ref{def-ULDP}, $(\mathcal{K}_S, \mathcal{Y}_P, \epsilon)$-ULDP maps sensitive data $v\in\mathcal{K}_S$ to only protected data set. Specifically, we can see from formula (\ref{uldp-1}) that no privacy protects are provided for non-sensitive data since each output in $\mathcal{Y}_I$ reveals the corresponding input in $\mathcal{K}_N$. Also, we can also find from formula (\ref{uldp-2}) that $(\mathcal{K}_S, \mathcal{Y}_P, \epsilon)$-ULDP  provides the same privacy protections as $\epsilon$-LDP for all sensitive data $v\in\mathcal{K}_S$.

\subsubsection{ID-LDP}

In ULDP, Murakami \textit{et al.} \cite{murakami2019utility} considered the sensitivity level of input data by directly separating the input data into sensitive data and non-sensitive data. However, Gu \textit{et al.} \cite{gu2020providing} further indicated that different data have distinct sensitivity levels. Thus, they presented the Input-Discriminative LDP (ID-LDP) which is a more fine-grained version of LDP. The notion of ID-LDP is defined as follows.

\begin{defn}[$\mathcal{E}$-ID-LDP]\label{def-ID-ldp}
	For a given privacy budget set $\mathcal{E}=\{\epsilon_v\}_{v\in\mathcal{K}}$, a randomized mechanism $\mathcal{M}$ satisfies $\mathcal{E}$-ID-LDP if and only if for any pairs of input values $v$ and $v'$, and for any possible output $y\in\mathcal{Y}$, it holds
	\begin{align}
	\mathbb{P}[\mathcal{M}(v) = y] \leq e^{ \textsf{r}(\epsilon_v, \epsilon_{v'})} \cdot \mathbb{P}[\mathcal{M}(v') = y]
	\end{align}
	where $ \textsf{r}(\cdot,\cdot)$ is a function of two privacy budget.
\end{defn}

It can be seen from Definition \ref{def-ID-ldp} that ID-LDP introduces the function  $ \textsf{r}(\epsilon_v, \epsilon_{v'})$ to quantify the indistinguishability between input values $v$ and $v'$ that have different privacy levels with privacy budget $\epsilon_v$ and $\epsilon{v'}$. The work in \cite{murakami2019utility} mainly considers the minimum function between $\epsilon_v$ and $\epsilon{v'}$ and formalizes the MinID-LDP as follows.

\begin{defn}[MinID-LDP]
	A randomized mechanism $\mathcal{M}$ satisfies $\mathcal{E}$-MinID-LDP if and only if it satisfies $\mathcal{E}$-ID-LDP with $ \textsf{r}(\epsilon_v, \epsilon_{v'})=\min\{\epsilon_v, \epsilon_{v'}\}$.
\end{defn}

That is, MinID-LDP always guarantees the worse-case privacy for the pair. Thus, MinID-LDP ensures better data utility by providing distinct protection for different inputs than standard LDP that provides the worse-case privacy for all data.

\subsubsection{PBP}

In addition, Takagi \textit{et al.} \cite{takagi2020poster} pointed that data providers can naturally choose and keep their privacy parameters secret since LDP perturbations occur in device side. Thus, they proposed a new privacy model Parameter Blending Privacy (PBP) as a generalization of standard LDP. PBP can not only keep the privacy parameters secret, but only improves the data utility through privacy amplification.

Let $\Theta$ be the domain of the privacy parameter.  Given a privacy budget $\theta \in \Theta$, let $\mathbb{P}(\theta)$ be the ratio of the number of times that $\theta$ is chosen to the number of users. Then, PBP is defined as follows.

\begin{defn}[\textsf{r}-PBP]\label{def-pbp}
	A randomized mechanism $\mathcal{M}$ satisfies \textsf{r}-PBP iff $\forall \theta\in\Theta, v,v'\in\mathcal{K}, y\in\mathcal{Y}$, $\exists\theta'\in\Theta$, it holds
	\begin{align}
		\mathbb{P}(\theta)\mathbb{P}[\mathcal{M}(v;\theta)=y] \leq e^{\textsf{r}(\theta)}\cdot \mathbb{P}(\theta')\mathbb{P}[\mathcal{M}(v';\theta')=y]
	\end{align}
	where the privacy function $\textsf{r}()$ returns a real number that denotes the strength of privacy protection.
\end{defn}

\textit{Comparisons and discussions.} Table~\ref{summary-ldp-variant} briefly summaries the various LDP variants from different perspectives. With various purposes, these variants extend the standard LDP into more generalized or granular versions based on different design ideas. Meanwhile, the main protocols for achieving such LDP variants are also proposed. Nonetheless, there are still some issues in new privacy notions remaining unsolved. For example, PBP only focuses on the privacy parameters that are chosen at the user-level. In other words, the correlations between data and privacy parameters are neglected in PBP. Similarly, ULDP can't be directly applied to scenarios that sensitive data and non-sensitive are correlated. Besides, MinID-LDP considers the minimum function to decide the privacy budget. There might be other functions that can provide better data utility.

\begin{table*}[t]
	\renewcommand\arraystretch{1} %行距
	\centering
	\caption{Summary of LDP variants (LDP is also listed for reference).}\label{summary-ldp-variant}
	\setlength{\tabcolsep}{0.5mm}{
		\fontsize{8pt}{\baselineskip}\selectfont
		\begin{threeparttable}
			\begin{tabular}{c|c|c|c|c|c|c}
				\bottomrule[1pt]
				\textbf{LDP Variants} &\textbf{Definition} &\textbf{Purpose} &\textbf{Design Idea}  &\textbf{Target Data Type}  &\textbf{Main Protocol}  &\textbf{$=$ LDP?}\\ \hline
				LDP \cite{wang2017protocol}    &$\frac{\mathbb{P}[\mathcal{M}(v)=y]}{\mathbb{P}[\mathcal{M}(v')=y]}\leq e^\epsilon$        &-             &-              &All data type         &\tabincell{c}{RR-based method}     &-\\ \hline
				\tabincell{c}{$(\epsilon,\delta)$-LDP\\\cite{bassily2015local,wt2019locally}}  &See formula (\ref{eqn-eps-delta-ldp})    &\tabincell{c}{A relaxed variant\\of LDP}    &\tabincell{c}{LDP fails with a small\\probability $\delta$} &All data type &\tabincell{c}{RR-based method}  &When $\delta =0$   \\\hline
				\tabincell{c}{BLENDER\\\cite{avent2017blender}} &same as $(\epsilon,\delta)$-DP &\tabincell{c}{Improve data utility\\by combine global\\DP and LDP} &\tabincell{c}{Group user pool} &Categorical data  &Laplace mechanism  &-  \\\hline
				\tabincell{c}{Local $d$-privacy\\\cite{alvim2018metric} } &$\frac{\mathbb{P}[\mathcal{M}(v)=y]}{\mathbb{P}[\mathcal{M}(v')=y]}\leq e^{\epsilon\cdot \textsf{d}(v,v')}$       &\tabincell{c}{Enhance data utility\\for metric spaces } &Metric-based method  &\tabincell{c}{Metric data,\\e.g., location data } &\tabincell{c}{Discrete Laplace\\Geometric\\mechanisms} &- \\\hline
				CLDP \cite{gursoy2019secure}  &$\frac{\mathbb{P}[\mathcal{M}(v)=y]}{\mathbb{P}[\mathcal{M}(v')=y]}\leq e^{\alpha\cdot  \textsf{d}(v,v')}$         &\tabincell{c}{Solve the problem of a\\small number of users}  &Metric-based method &\tabincell{c}{Categorical data} &\tabincell{c}{Exponential\\mechanism}   &-   \\\hline
				PLDP \cite{nie2019utility}  &$\frac{\mathbb{P}[\mathcal{M}(v)=y]}{\mathbb{P}[\mathcal{M}(v')=y]}\leq e^{\epsilon_U}$ &\tabincell{c}{Achieve granular\\privacy constraints} &\tabincell{c}{Advanced\\combination \cite{nie2019utility}\\PCE \cite{chen2016private}} &Categorical data &\tabincell{c}{RR-based method}  &When $\epsilon_U = \epsilon$ \\\hline
				ULDP \cite{murakami2019utility} &See Definition \ref{def-ULDP}   &\tabincell{c}{Optimize\\data utility}  &\tabincell{c}{Only provide\\privacy guarantees\\for sensitive data} &Categorical data  &\tabincell{c}{RR-based method}  &\tabincell{c}{When $\mathcal{K}_S=\mathcal{K}$\\and $\mathcal{Y}_P=\mathcal{Y}$} \\\hline
				ID-LDP \cite{gu2020providing} &$\frac{\mathbb{P}[\mathcal{M}(v)=y]}{\mathbb{P}[\mathcal{M}(v')=y]}\leq e^{ \textsf{r}(\epsilon_x, \epsilon_{x'})}$  &\tabincell{c}{Provide\\input-discriminative\\protection for\\different inputs} &\tabincell{c}{Quantify\\indistinguishability} &Categorical data  &Unary Encoding  &\tabincell{c}{When $\epsilon_v =\epsilon$\\for each value $v$ } \\\hline
				PBP \cite{takagi2020poster} &See Definition \ref{def-pbp}   &\tabincell{c}{Achieve privacy\\amplification of LDP} &\tabincell{c}{Keep privacy\\parameters secret} &Categorical data &\tabincell{c}{RR-based method}  &-  \\\hline
				\toprule[1pt]			\end{tabular}
		\end{threeparttable}
	}
\end{table*}

\section{Frequency Estimation with LDP}\label{sec-fre}

This Section summarizes the state-of-the-art LDP algorithms for frequency estimation. Frequency estimation, which is equivalent to histogram estimation, aims at computing the frequency of each given value $v \in \mathcal{K}$, where $|\mathcal{K}|=k$. Besides, we further subdivide the frequency-based task under LDP into several more specific tasks. In what follows, we will introduce each LDP protocol in the view of \textsf{randomization}, \textsf{aggregation}, and \textsf{estimation}, as described in Section~\ref{sec-framework}.

Based on the Definition~\ref{defn-eps-ldp} of LDP, a more visual definition of LDP protocol \cite{wang2017protocol} can be given as follows.
\begin{defn}[$\epsilon$-LDP Protocol]\label{local-protocol}
Consider two probabilities $p>q$. A local protocol given by $\mathcal{M}$ such that a user reports the true value with $p$ and reports each of other values with $q$, will satisfy $\epsilon$-LDP if and only if it holds $p\leq q\cdot e^\epsilon$.
\end{defn}

Based on the Theorem 2 in \cite{wang2017protocol}, the variance for the noisy number of the value $v$ (i.e., $\widehat{N}_v$) among $N$ users will be $\text{Var}[\widehat{N}_v]=\frac{Nq(1-q)}{(p-q)^2}+\frac{N\widehat{f}_v(1-p-q)}{p-q}$, where $f_v$ is the frequency of the value $v\in\mathcal{K}$. Thus, the variance is
\begin{align}\label{var-ldp}
    \text{Var}[\widehat{f}_v]=\frac{q(1-q)}{N(p-q)^2}+\frac{\widehat{f}_v(1-p-q)}{N(p-q)}.
\end{align}
It can be seen that the variance of Eq.~(\ref{var-ldp}) will be dominated by the first term when $f_v$ is small. Hence, the approximation of the variance in Eq.~(\ref{var-ldp}) can be denoted as
\begin{align}\label{var-ldp-approx}
    \text{Var}^*[\widehat{f}_v]=\frac{q(1-q)}{N(p-q)^2}.
\end{align}
In addition, it holds that Var$^*$=Var when $p+q=1$.

\subsection{General Frequency Estimation on Categorical Data}\label{sec-fre-basic}

This section summarizes the general LDP protocols for frequency estimation on categorical data and shows the performance of each protocol. The encoding principle of the existing LDP protocols can be concluded as direct perturbation, unary encoding, hash encoding, transformation, and subset selection.

\subsubsection{Direct Perturbation}

The most basic building block for achieving LDP is direct perturbation that perturbs data directly by randomization.

\textbf{Binary Randomized Response (BRR) \cite{warner1965randomized,kairouz2014extremal}} is the basic randomized response technique that focuses on binary values, that is the cardinality of value domain is 2. Section~\ref{sec-randomized-response} has introduced the basic randomized response technique that focuses on binary values. Based on this, BRR is formally defined as follows.

\textsf{Randomization}. Each value $v$ is perturbed by
\begin{align}\label{prob-brr}
    \bp{\mathcal{M}(v)=v^*}=
    \begin{cases}
    p=\frac{e^\epsilon}{e^\epsilon +1},&\text{if~}v^*=v,\\
    q=\frac{1}{e^\epsilon +1},&\text{if~}v^*\neq v.
    \end{cases}
\end{align}

\textsf{Aggregation and Estimation}. Let $\widehat{N}_v$ be the total number of received value $v$ after aggregation. The estimated frequency $\widehat{f}_v$ of value $v$ can be computed as $\widehat{f}_v=\left(\frac{\widehat{N}_v}{N}-\frac{1}{e^\epsilon +1}\right)\cdot\frac{e^\epsilon +1}{e^\epsilon -1}$.

Observe that the probability that $v^*=v$ varies from $\frac{1}{e^\epsilon +1}$ to $\frac{e^\epsilon}{e^\epsilon +1}$. The ratio of the respective probabilities for different values of $v$ will be at most $e^\epsilon$. Therefore, BRR satisfies $\epsilon$-LDP. Based on Eq.~(\ref{var-ldp-approx}), the variance of BRR is
\begin{align}
    \text{Var}^*_{BRR}[\widehat{f}_v]= \frac{e^\epsilon}{N(e^\epsilon -1)^2}.
\end{align}

\textbf{Generalized Randomized Response (GRR) \cite{kairouz2016discrete,wang2016private}} extends the BRR to the case where the cardinality of total values is more than 2, that is $k>2$. GRR is also called Direct Encoding (DE) in \cite{wang2017protocol} or $k$-RR in \cite{kairouz2016discrete}. The process of GRR is given as follows.

\textsf{Randomization}. Each value $v$ is perturbed by
\begin{align}\label{prob-grr}
    \bp{\mathcal{M}(v)=v^*}=
    \begin{cases}
    p=\frac{e^\epsilon}{e^\epsilon +k-1},&\text{if~}v^*=v,\\
    q=\frac{1}{e^\epsilon +k-1},&\text{if~}v^*\neq v.
    \end{cases}
\end{align}

\textsf{Aggregation and Estimation}. Let $\widehat{N}_v$ be the total number of received value $v$ after aggregation. The estimated frequency $\widehat{f}_v$ of value $v$ can be computed as
\begin{align}
    \widehat{f}_v=\left(\frac{\widehat{N}_v}{N}-\frac{1}{e^\epsilon +k-1}\right)\cdot\frac{e^\epsilon +k-1}{e^\epsilon -1}.
\end{align}

Observe that the probability that $v^*=v$ varies from $\frac{1}{e^\epsilon +k-1}$ to $\frac{e^\epsilon}{e^\epsilon +k-1}$. The ratio of the respective probabilities for different values of $v$ will be at most $e^\epsilon$. Therefore, GRR satisfies $\epsilon$-LDP. Based on Eq.~(\ref{var-ldp-approx}), the variance of GRR is
\begin{align}
    \text{Var}^*_{GRR}[\widehat{f}_v]=\frac{(e^\epsilon +k-2)}{N(e^\epsilon -1)^2}.
\end{align}

\subsubsection{Unary Encoding}

Instead of perturbing the original value, we can perturb each bit of a vector that is generated by encoding the original value $v$. This method is called \textbf{Unary Encoding (UE)} \cite{wang2017protocol} that is achieved as follows.

\textsf{Randomization}. UE encodes each value $v \in \mathcal{K}$ into a binary bit vector $B$ with size $k$, where the $v$-th bit is 1, that is $B=[0,\cdots,0,1,0,\cdots,0]$. Each bit of $B$ is perturbed by
\begin{align}\label{prob-ue}
    \bp{B^*[i]=1}=
    \begin{cases}
    p,&\text{~if~}B[i]=1, \\
    q,&\text{~if~}B[i]=0,
    \end{cases}
\end{align}
where $p>q$.

\textsf{Aggregation and Estimation}. Assume the number of ones in the $v$-th bit among all original $N$ vectors and all received $N$ vectors are $N^1_v$ and $\bar{N}^1_v$, respectively. Based on Eq.~(\ref{prob-ue}), we have $\bar{N}^1_v = N^1_vp+(N-N^1_v)q$. Thus, the estimated number of value $v$ is $\widehat{N}^1_v = \frac{\bar{N}^1_v-Nq}{p-q}$. Then, the frequency $f_v$ of the value $v$ is computed as
\begin{align}\label{fre-ue}
    \widehat{f}_v=\frac{\widehat{N}^1_v}{N}=\left( \frac{\bar{N}^1_v}{N} - q \right )/(p-q).
\end{align}

Based on Eq.~(\ref{prob-ue}), for any inputs $v_1\in \mathcal{K}$ and $v_2\in \mathcal{K}$, and the output $B^*$, it holds that
\begin{align}
	\frac{\bp{B^*|v_1}}{\bp{B^*|v_2}}
	&=\frac{\prod_{i\in[k]}\bp{B^*[i]|v_1}}{\prod_{i\in[k]}\bp{B^*[i]|v_2}}\nonumber\\
	&\leq \frac{\mathbb{P}[B^*[v_1]=1|v_1]\mathbb{P}[B^*[v_2]=0|v_1]}{\mathbb{P}[B^*[v_1]=1|v_2]\mathbb{P}[B^*[v_2]=0|v_2]}\label{proof-ue}
\end{align}
where ``$\leq$'' is achieved since the bit vectors differ only in positions $v_1$ and $v_2$. There are four cases when choosing values for positions $v_1$ and $v_2$. That is,
\begin{align}
	\begin{cases}
	\textcircled{1}\frac{\bp{B^*[v_1]=0|v_1}\cdot \bp{B^*[v_2]=0|v_1}}{\bp{B^*[v_1]=0|v_2}\cdot \bp{B^*[v_2]=0|v_2} }\vspace{1mm}\\
	\textcircled{2}\frac{\bp{B^*[v_1]=0|v_1}\cdot \bp{B^*[v_2]=1|v_1}}{\bp{B^*[v_1]=0|v_2}\cdot \bp{B^*[v_2]=1|v_2}}\vspace{1mm}\\
	\textcircled{3}\frac{\bp{B^*[v_1]=1|v_1}\cdot \bp{B^*[v_2]=0|v_1}}{\bp{B^*[v_1]=1|v_2}\cdot \bp{B^*[v_2]=0|v_2}}\vspace{1mm}\\
	\textcircled{4}\frac{\bp{B^*[v_1]=1|v_1}\cdot \bp{B^*[v_2]=1|v_1}}{\bp{B^*[v_1]=1|v_2}\cdot \bp{B^*[v_2]=1|v_2}}
	\end{cases}
\end{align}
It can be verified that a vector with position $v_1$ being 1 and position $v_2$ being 0 will maximize the ratio (i.e., the case \textcircled{3}).

Based on Eq.~(\ref{prob-ue}), UE satisfies $\epsilon$-LDP if and only if it follows that
\begin{align}
	\frac{\bp{B^*[v_1]=1|v_1}\cdot \bp{B^*[v_2]=0|v_1}}{\bp{B^*[v_1]=1|v_2}\cdot \bp{B^*[v_2]=0|v_2}}=\frac{p(1-q)}{q(1-p)}\leq e^\epsilon \label{ue}
\end{align}
Therefore, letting the equal sign in Eq.~(\ref{ue}) hold, we can set $p$ as follows:
\begin{align}\label{ue-pro}
 p = \frac{qe^\epsilon}{1-q+qe^\epsilon}.
\end{align}
Applying Eq.~(\ref{ue-pro}) to Eq.~(\ref{var-ldp}), the variance of UE is
\begin{align}\label{var-ue}
    \text{Var}^*_{\text{UE}}[\widehat{f}_v]=\frac{(1-q+qe^\epsilon)^2}{Nq(1-q)(e^\epsilon-1)^2}.
\end{align}

\textbf{Symmetric UE (SUE) \cite{wang2017protocol}} is the symmetric version of UE when choosing $p$ and $q$ such that $p+q=1$. Based on this observation and Eq.~(\ref{ue-pro}), we can derive $p=\frac{e^{\epsilon/2}}{e^{\epsilon/2} + 1}$ and $q=\frac{1}{e^{\epsilon/2} + 1}$. Then, the frequency can be computed based on Eq.~(\ref{fre-ue}).  And the variance of SUE is
\begin{align}
    \text{Var}^*_{\text{SUE}}[\widehat{f}_v]=\frac{e^{\epsilon/2}}{N(e^{\epsilon/2}-1)^2}.
\end{align}

\textbf{Optimized UE (OUE) \cite{wang2017protocol}} is to minimize the Eq.~(\ref{var-ue}). By making the partial derivative of Eq.~(\ref{var-ue}) with respect to $q$ equals to 0, we can get the formula $\frac{1}{(e^\epsilon-1)^2}\left(\frac{e^{2\epsilon}}{(1-q)^2}-\frac{1}{q^2}\right)$. By solving this, we can obtain
\begin{align}\label{oue-pq}
    p=\frac{1}{2},~q=\frac{1}{e^\epsilon+1}.
\end{align}
The estimated frequency can be computed by Eq.~(\ref{fre-ue}). By combining the Eqs.~(\ref{var-ue}) and (\ref{oue-pq}), the variance of OUE is
\begin{align}\label{var-oue}
    \text{Var}^*_{\text{OUE}}[\widehat{f}_v]=\frac{4e^\epsilon}{N(e^\epsilon-1)^2}.
\end{align}

\subsubsection{Hash Encoding}

In the same way of UE, Basic RAPPOR \cite{erlingsson2014rappor} encodes each value $v \in \mathcal{K}$ into a length-$k$ binary bit vector $B$ and conducts \textsf{Randomization} with the following two steps.

\textit{Step 1: Permanent randomized response}: Generate $B_1$ with the probability
\begin{align}\label{RAPPOR-step1}
    \bp{B_1[v]=1}=
    \begin{cases}
    1-\frac{1}{2}r, &\text{~if~}B[v]=1,\\
    \frac{1}{2}r, &\text{~if~}B[v]=0.
    \end{cases}
\end{align}
where $r$ is a user-tunable parameter that controls the level of longitudinal privacy guarantee.

\textit{Step 2: Instantaneous randomized response}:
Perturb $B_1$ with the following probability distribution (i.e., UE)
\begin{align}\label{RAPPOR-step2}
    \bp{B^*[i]=1}=
    \begin{cases}
    p,&\text{~if~}B_1[i]=1, \\
    q,&\text{~if~}B_1[i]=0.
    \end{cases}
\end{align}

From the proof in \cite{erlingsson2014rappor}, the \textit{Permanent randomized response} (i.e., \textit{Step 1}) achieves $\epsilon$-LDP for $\epsilon=2\ln{\frac{1-r/2}{r/2}}$. The communication and computing cost of Basic RAPPOR is $\Theta (k)$ for each user, and $\Theta (Nk)$ for the aggregator. However, Basic RAPPOR doesn't scale to the cardinality $k$.

\textbf{RAPPOR \cite{erlingsson2014rappor}} adopts Bloom filters \cite{bloom1970space} to encode each single element based on a set of $m$ hash functions $\mathbb{H}=\{H_1, H_2, \cdots,H_m\}$. Each hash function firstly outputs an integer in $\{0,1,\cdots,k-1\}$. Then, each value $v$ is encoded as a $k$-bit binary vector $B$ by
\begin{align}
    B[i]=
    \begin{cases}
    1,&\text{if~}\exists H\in\mathbb{H},s.t.,H(v)=i,\\
    0,&\text{otherwise}.
    \end{cases}
\end{align}
Next, RAPPOR uses the same processes (i.e., Eqs.~(\ref{RAPPOR-step1}) and (\ref{RAPPOR-step2})) as Basic RAPPOR to conduct randomization.

From the proof in \cite{erlingsson2014rappor}, RAPPOR achieves $\epsilon$-LDP for $\epsilon=2m\ln{\frac{1-r/2}{r/2}}$. Moreover, the communication cost of RAPPOR is $\Theta (k)$ for each user. However, the computation cost of the aggregator in RAPPOR is higher than Basic RAPPOR due to the LASSO regression.

\textbf{O-RAPPOR \cite{kairouz2016discrete}} is proposed to address the problem of holding no prior knowledge about the attribute domain. Kairouz~\textit{et al.}~\cite{kairouz2016discrete} examined discrete distribution estimation when the open alphabets of categorical attributes are not enumerable in advance. They applied hash functions to map the underlying values at first. Then, the hashed values will be involved in a perturbation process, which is independent of the original values. On the basis of RAPPOR, Kairouz~\textit{et al.} adopted the idea of hash cohorts. Each user $u_i$ will be assigned to a cohort $c_i$ that is sampled i.i.d. from a uniform distribution over $\mathcal{C}=\{1,\cdots,C\}$. Each $c\in\mathcal{C}$ provides an independent view of the underlying distribution of strings. Based on hash cohorts, O-RAPPOR applies hash functions on a value $v$ in cohort $c$ before using RAPPOR and generates an independent $h$-bit hash Bloom filter $BLOOM_c$ for each cohort $c$, where the $j$-th bit of $BLOOM_c$ is $1$ if $HASH_{c,h'}(v)=j$ for any $h'\in[1\cdots h]$. Next, the perturbation on $BLOOM_c$ follows the same strategy in RAPPOR.

\textbf{O-RR \cite{kairouz2016discrete}} is proposed to deal with non-binary attributes. It integrates hash cohorts into $k$-RR to deal with the case where the domain of attribute is unknown. Users in a cohort use their cohort hash function to project the value space into $k$ disjoint subsets, i.e., $x_i=HASH_{c}(v)\mod k = HASH_{c}^k(v)$. Next, the O-RR perturbs the input value $v$ as follows:
\begin{align}\label{prob-o-rr}
    \bp{v^*|v}=\frac{1}{C(e^\epsilon+k-1)}
    \begin{cases}
        e^\epsilon, &\text{if~}HASH_c^k(v)=v^*,\\
        1, &\text{if~}HASH_c^k(v)\neq v^*.
    \end{cases}
\end{align}
Note that Eq.~(\ref{prob-o-rr}) contains a factor of $C$ compared to Eq.~(\ref{prob-grr}). This is because each value $v$ belongs to one of the cohorts. The error bound of O-RR is the same as $k$-RR, but incurs more time cost due to hash and cohort operations.

To reduce communication and computation cost, local hashing (LH) \cite{wang2017protocol} is proposed to hash the input value into a domain $[g]$ such that $g<k$. Denote $\mathbb{H}$ as the universal hash function family. Each input value is hashed into a value in $[g]$ by hash function $H\in\mathbb{H}$. The universal property requires that
\begin{align}\label{lh-condition}
\forall v_1,v_2\in[k],v_1\neq v_2:\underset{H\in\mathbb{H}}P[H(v_1)=H(v_2)]\leq \frac{1}{g}.
\end{align}

\textsf{Randomization}. Given any input value $v\in[k]$, LH first outputs a value $x$ in $[g]$ by hashing, that is $x=H(v)$. Then, LH perturbs $x$ with the following distribution
\begin{align}\label{prob-lh}
\forall i\in[g], \bp{y=i}=
\begin{cases}
p=\frac{e^\epsilon}{e^\epsilon+g-1},&\text{~if~}x=i,\\
q=\frac{1}{e^\epsilon+g-1},&\text{~if~}x\neq i.
\end{cases}
\end{align}
After perturbation, each user sends $\left \langle H,y \right \rangle$ to the aggregator. Based on Eq.~(\ref{prob-lh}), we can know that LH satisfies $\epsilon$-LDP since it always holds that $p\leq qe^\epsilon$.

\textsf{Aggregation and Estimation}. Assume we aim to estimate the frequency $f_v$ of the value $v$. The aggregator counts the total number that $\left \langle H,y \right \rangle$ supports value $v$, denoted as $\theta$. That is, for each report $\left \langle H,y \right \rangle$, if it holds that $H(v)=y$, then $\theta = \theta +1$. Based on Eq.~(\ref{prob-lh}), it holds that
\begin{align}\label{real-pro-lh}
p^*=p,~q^*=\frac{1}{g}p+\frac{g-1}{g}q=\frac{1}{g},
\end{align}
where $p^*$ is the probability of keeping unchanged of an input value and $q^*$ is the probability of flipping an input value.

Then, while aggregating in the server, we have
\begin{align}\label{aggregation-lh}
f_vp^*+(1-f_v)q^*=\theta/N.
\end{align}
Based on Eqs.~(\ref{real-pro-lh}) and (\ref{aggregation-lh}), we can get the estimated frequency of the value $v$, that is
\begin{align}\label{estimation-lh}
\widehat{f}_v = \left(\frac{g\theta}{N} - 1\right)\cdot\frac{e^\epsilon+g-1}{ge^\epsilon-e^\epsilon -g +1}.
\end{align}
By taking $p=p^*$, $q=q^*$ into Eq.~(\ref{var-ldp}), the variance of LH is
\begin{align}\label{var-lh}
\text{Var}^*_{\text{LH}}[\widehat{f}_v]=\frac{(e^\epsilon+g-1)^2}{N(g-1)(e^\epsilon-1)^2}.
\end{align}

Local hashing will become \textbf{Binary Local Hashing (BLH) \cite{wang2017protocol}} when $g=2$. In BLH, each hash function $H\in\mathbb{H}$ hashes an input from $[k]$ into one bit. 

%Based on Eq.~(\ref{lh-condition}), the universal property requires that
%\begin{align}\label{blh-condition}
%\forall v_1,v_2\in[k],v_1\neq v_2:\underset{H\in\mathbb{H}}P[H(v_1)=H(v_2)]\leq \frac{1}{2}.
%\end{align}

\textsf{Randomization}. Based on Eq.~(\ref{prob-lh}), the randomization of BLH follows the probability distribution as
\begin{align}\label{prob-blh}
\bp{y=1}=
\begin{cases}
p=\frac{e^\epsilon}{e^\epsilon+1},&\text{~if~}x=1,\\
q=\frac{1}{e^\epsilon+1},&\text{~if~}x=0.
\end{cases}
\end{align}

\textsf{Aggregation and Estimation}. Based on LH, it holds that
$p^*=p$ and $q^*=\frac{1}{2}p+\frac{1}{2}q=\frac{1}{2}$.
When the reported supports of value $v$ is $\theta$, based on Eq.~(\ref{estimation-lh}). the estimated frequency can be computed as
\begin{align}
\widehat{f}_v = \left(\frac{2\theta}{N} - 1\right)\cdot\frac{e^\epsilon+1}{e^\epsilon -1}.
\end{align}
And the variance of BLH is
\begin{align}\label{var-blh}
\text{Var}^*_{\text{BLH}}[\widehat{f}_v]=\frac{(e^\epsilon+1)^2}{N(e^\epsilon-1)^2}.
\end{align}

\textbf{Optimized LH (OLH) \cite{wang2017protocol}} aims to choose an optimized $g$ to compromise the information losses between hash step and randomization step. Based on Eq.~(\ref{var-lh}), we can minimize the variance of LH by making the partial derivative of Eq.~(\ref{var-lh}) with respect to $g$ equals to 0. That is, it is equivalent to solve the following equation $(e^\epsilon-1)^2\cdot g - (e^\epsilon-1)^2(e^\epsilon +1) = 0$. By solving it, the optimal $g$ is $g=e^\epsilon +1$, where $g=\left \lfloor e^\epsilon +1 \right \rfloor$ in practice. When the reported supports of value $v$ is $\theta$, based on Eq.~(\ref{estimation-lh}), the estimated frequency is $\widehat{f}_v = \frac{2(g\theta-N)}{N(e^\epsilon -1)}$. And, the variance of OLH is
\begin{align}\label{var-olh}
\text{Var}^*_{\text{OLH}}[\widehat{f}_v]=\frac{4e^\epsilon}{N(e^\epsilon-1)^2}.
\end{align}

\subsubsection{Transformation}

The transformation-based method is usually adopted to reduce the communication cost.

\textbf{S-Hist \cite{bassily2015local}} is proposed to produce a succinct histogram that contains the most frequent items (i.e., ``heavy hitters''). Bassily and Smith \cite{bassily2015local} have proved that S-Hist achieves asymptotically optimal accuracy for succinct histogram estimation. S-Hist randomly selects only one bit from the encoded vector based on random matrix projection, which reduces the communication cost. The specific process of S-Hist is as follows, which includes an additional initialization step.

\textsf{Initialization}. The aggregator generates a random projection matrix $\Phi \in \{-\frac{1}{\sqrt{b}}, \frac{1}{\sqrt{b}}\}^{b\times k}$, where each element of $\Phi$ is extracted from the set $\{-\frac{1}{\sqrt{b}}, \frac{1}{\sqrt{b}}\}$. The magnitude of each column vector in $\Phi$ is 1, and the inner product of any two different column vectors is 0. Here $b$ is a constant parameter determined by error bound, where error is defined as the maximum distance between the estimated and true frequencies, that is $\text{max}_{v\in\mathcal{K}}|\widehat{f}_v - f_v|$.

\textsf{Randomization}. Assume the input value $v$ is the $v$-th element of domain $\mathcal{K}$. We encode $v$ as $Encode(v)=\left \langle j, x \right \rangle$, where $j$ is chosen uniformly at random from $[b]$, and $x$ is the $v$-th element of the $j$-th row of $\Phi$, that is, $x= \Phi[j,v]$. Then, we randomize $x$ as follows:
\begin{align}\label{prob-S-Hist}
    z=
    \begin{cases}
    c_\epsilon bx, &\text{w.p.~}\frac{e^\epsilon}{e^\epsilon +1},\\
    -c_\epsilon bx, &\text{w.p.~}\frac{1}{e^\epsilon +1},
    \end{cases}
\end{align}
where $c_\epsilon=\frac{e^\epsilon +1}{e^\epsilon -1}$.
After perturbation, each user $u_i(i\in[1,N])$ sends $\left \langle j_i, z_i \right \rangle$ to the aggregator.

\textsf{Aggregation and Estimation}. Upon receiving the report $\left \langle j^i, z^i \right \rangle$ of each user $u_i$, the estimation for the $v$-th element of $\mathcal{K}$ is computed by
\begin{align}
    \widehat{f}_v=\sum_{i\in[1,N]}z^i\cdot \Phi[j^i,v].
\end{align}

Based on Eq. (\ref{prob-S-Hist}), it's easy to know that S-Hist satisfies $\epsilon$-LDP for every choice of the index $j$. What's more, Bassily and Smith \cite{bassily2015local} proved that the $L_\infty$-error of S-Hist is bounded by $O\left(\frac{1}{\epsilon}\sqrt{\frac{\log (k/\beta)}{N}}\right)$ with probability as least $1-\beta$.

\textbf{Hadamard Randomized Response (HRR) \cite{apple2017local, cormode2018marginal, acharya2019hadamard}} is a useful tool to handle sparsity by transforming the information contained in sparse vectors into a different orthonormal basis. HRR adopts Hadamard transformation (HT) to handle the situation where the inputs and marginals of individual users are sparse. HT is also called discrete Fourier transform, which is described by an orthogonal and symmetric matrix $\phi $ with dimension $2^k\times 2^k$. Each row/column in $\phi $ is denoted as $\phi_{i,j}=2^{-k/2}(-1)^{\left \langle i,j \right \rangle}$, where $\left \langle i,j \right \rangle$ denotes the number of 1's that $i$ and $j$ agree on in their binary representation. When a value $v_i$ is presented as a sparse binary vector $B_i$, the full Hadamard transformation of the input is the $B_i$-th column of $\phi $, that is, the Hadamard coefficient $o_i=\phi \times B_i$.

\textsf{Randomization}. User $i$ samples an index $j\in 2^k$ and perturbs $\phi_{B_i,j} \in \{-1,1\}$ by using BRR that keeps true value with probability $p$ and flips the value with probability $1-p$. Then, the user $i$ reports the perturbed coefficient $\widehat{o}_i$ and the index $j$ to the aggregator. As we can see, the communication cost is $O(\log k +1)=O(\log k)$.

\textsf{Aggregation and Estimation}. Assume the observed sum of all received perturbed coefficient with index $j$ is $O_j$. Then, the unbiased estimation of the $j$-th Hadamard coefficient $\widehat{o}_j$ (with the $2^{-k/2}$ factor scaled) is computed by
\begin{align}
\widehat{o}_j = \frac{O_j}{2^{k/2}(2p-1)}.
\end{align}
In this way, the aggregator can compute the unbiased estimations of all coefficients and apply inverse transformation to produce the final frequency estimation $\widehat{f}$. 

Based on the proof in \cite{cormode2018marginal, cormode2019answering}, the variance of HRR is
\begin{align}
\text{Var}_{\text{HRR}}[\widehat{f}]=\frac{4p(1-p)}{N(2p-1)^2}.
\end{align}
By setting $p=\frac{e^\epsilon}{e^\epsilon +1}$ to ensure LDP, the variance is $\frac{4e^\epsilon}{N(e^\epsilon-1)^2}$. Thus, HRR provides a good compromise between accuracy and communication cost. Besides, the computation overhead in the aggregator is $O(N+k\log k)$, versus $O(Nk)$ for OLH.

Furthermore, Jayadev \textit{et al.}~\cite{acharya2019hadamard} designed a general family of LDP schemes. Based on Hadamard matrices, they choose the optimal privatization scheme from the family for high privacy with less communication cost and higher efficiency.

\begin{algorithm}[t]\small
	\setstretch{1.2} 
	\caption{\small{The Randomization of $\omega $-SM}}\label{algo-m-sm}
	\KwIn{Input $v\in \mathcal{K}$, privacy budget $\epsilon$, subset size $\omega $}
	\KwOut{Subset $z \subseteq \mathcal{K}$ with size $\omega $}
	Initialize $z$ as an empty set, $z=\{\}$\;
	Sample a Bernoulli variable $u$ with $\bp{u=1}=\frac{\omega e^\epsilon}{\omega e^\epsilon +k -\omega }$\;
	\eIf{$u=1$}
	{Insert $v$ into $z$, $z=z\cup \{v\}$\;
		Randomly sample $\omega -1$ elements $Y$ from $\mathcal{K} - \{v\}$\;
		Add elements in $Y$ to $z$, $z=z\cup Y$\;}
	{Randomly sample $\omega $ elements $Y$ from $\mathcal{K} - \{v\}$\;
		Add elements in $Y$ to $z$, $z=z\cup Y$\;}
	\textbf{return} $z$\; 
\end{algorithm}

\subsubsection{Subset Selection}

The main idea of subset selection is randomly select $\omega $ items from the domain $\mathcal{K}$.

\textbf{$\omega $-Subset Mechanism ($\omega$-SM) \cite{wang2016mutual, WangHNZWXY19}} is proposed to randomly reports a subset $z$ with size $\omega $ of the original attribute domain $\mathcal{K}$, that is $z \subseteq \mathcal{K}$. Essentially, the output space $z$ is the power set of the data domain $\mathcal{K}$. And the conditional probabilities of any input $v \in \mathcal{K}$, output $z \subseteq \mathcal{K}$ are as follows:
\begin{align}\label{prob-se}
    \bp{z|v}=
    \begin{cases}
    \frac{\omega e^\epsilon}{\omega e^\epsilon +k -\omega }/\binom{k}{\omega }, &\text{if~} |z|=\omega  \text{~and~}v \in z,\\
    \frac{\omega }{\omega e^\epsilon +k -\omega }/\binom{k}{\omega }, &\text{if~} |z|=\omega  \text{~and~} v \notin z,\\
    0, &\text{if~} |z|\neq \omega .
    \end{cases}
\end{align}

As we can see, when $\omega =1$, the $1$-SM is equivalent to generalized randomized response (GRR) mechanism. 

\textsf{Randomization}. Based on Eq.~(\ref{prob-se}), the randomization procedure of $\omega $-SM is shown in Algorithm~\ref{algo-m-sm}. Observe that the core part of randomization is randomly sampling $\omega -1$ or $\omega $ elements from $\mathcal{K} - \{v\}$ without replacement. 

\textsf{Aggregation and Estimation}. Denote $f_i$ and $\bar{f}_i$ as the real and the received frequency of the $i$-th value $v_i$, respectively. Upon receiving a private view $z_i$, it will increase $\bar{f}_i$ for each $v_i \in z_i$. Based on Algorithm~\ref{algo-m-sm}, we can know that the true positive rate $\theta_{tr}$ is $\theta_{tr} = \frac{\omega e^\epsilon}{\omega e^\epsilon +k -\omega }$, which is the probability of $v_i$ staying unchanged when the input value is $v_i$. And the false positive rate $\theta_{fr}$ is $\theta_{fr} = \frac{\omega e^\epsilon}{\omega e^\epsilon +k -\omega }\cdot \frac{\omega -1}{k-1} +
\frac{k-\omega }{\omega e^\epsilon +k -\omega }\cdot \frac{\omega }{k-1}$, which is the probability of $v_{i'}$ showing in the private view when the input value is $v_i~(i\neq i')$. Therefore, the expectation of $\bar{f}_i$ is $\mathbb{E}[\bar{f}_i]=f_i\cdot \theta_{tr} + (1-f_i)\cdot \theta_{fr}$.

Thus, we can get the estimated frequency of $\widehat{f}_i$ is
\begin{align}
    \widehat{f}_i = \frac{\bar{f}_i - \theta_{fr}}{\theta_{tr} - \theta_{fr}}.
\end{align}

\textit{Comparisons}. Table~\ref{compare-ldp-fre} summarizes the general LDP protocols from the perspective of encoding principles. The error bound is measured by $L_\infty$-norm. BRR and GRR are direct perturbation-based methods, which are suitable for low-dimensional data. BRR has a communication cost of $O(\log 2)=O(1)$ and has a smaller error bound than other mechanisms. GRR is the general version of BRR when $k>2$, of which the communication cost and error bound are both sensitive to domain size $k$. Both SUE and OUE are unary encoding-based methods. They have the same communication cost and error bound. RAPPOR, O-RAPPOR, O-RR, BLH, and OLH are hash encoding-based methods. RAPPOR, O-RAPPOR, and O-RR have larger error bounds and are relatively harder to implement since they involve Bloom filters and hash cohorts. BLH and OLH have smaller error bounds and are applicable to all privacy regime and any attribute domain. S-Hist and HRR are transformation-based methods, which have a lower communication cost. $\omega$-SM is a subset selection-based method, which reduces the communication cost.

\begin{table*}[htb]
	\renewcommand\arraystretch{1} %行距
	\centering
	\caption{Comparisons of general LDP protocols for frequency estimation}\label{compare-ldp-fre}
	\setlength{\tabcolsep}{2mm}{
		\fontsize{7pt}{\baselineskip}\selectfont
		\begin{threeparttable}
			\begin{tabular}{c|c|c|c|c|c}
				\bottomrule[1pt]
				\textbf{\tabincell{c}{Encoding Principle}} &\textbf{LDP Algo.}  &\textbf{\tabincell{c}{Comm. Cost}} &\textbf{Error Bound} &\textbf{Variance} &\textbf{\tabincell{c}{Know Domain?}} \\ \hline
				\multirow{2}{*}{\textbf{\tabincell{l}{Direct Perturbation}}}  &BRR \cite{warner1965randomized,kairouz2014extremal} &$O(1)$  &$O(\frac{1}{\epsilon\sqrt{N}})$  &$\frac{e^\epsilon}{N(e^\epsilon -1)^2}$  &Y \\ \cline{2-6}
				&\tabincell{c}{GRR \cite{kairouz2016discrete} (or, DE/$k$-RR)}  &$O(\log k)$  &$O(\frac{\sqrt{k\log{k}}}{\epsilon\sqrt{N}})$  &$\frac{e^\epsilon +k-2}{N(e^\epsilon -1)^2}$  &Y \\ \hline
				\multirow{2}{*}{\textbf{\tabincell{l}{Unary Encoding}}} &SUE \cite{wang2017protocol}  &$O(k)$  &$O(\frac{\sqrt{\log{k}}}{\epsilon\sqrt{N}})$   &$\frac{e^{\epsilon/2}}{N(e^{\epsilon/2}-1)^2}$  &Y \\ \cline{2-6}
				&OUE \cite{wang2017protocol} &$O(k)$  &$O(\frac{\sqrt{\log{k}}}{\epsilon\sqrt{N}})$ &$\frac{4e^\epsilon}{N(e^\epsilon-1)^2}$  &Y \\ \hline
				\multirow{5}{*}{\textbf{\tabincell{l}{Hash Encoding}}}  &RAPPOR \cite{erlingsson2014rappor} &$\Theta(k)$  &$O(\frac{k}{\epsilon\sqrt{N}})$   &$\frac{e^{\epsilon/2}}{N(e^{\epsilon/2}-1)^2}$   &Y \\ \cline{2-6}
				&O-RAPPOR \cite{kairouz2016discrete} &$\Theta(k)$  &$O(\frac{k}{\epsilon\sqrt{N}})$  &$\frac{e^{\epsilon/2}}{N(e^{\epsilon/2}-1)^2}$  &N \\ \cline{2-6}
				&O-RR \cite{kairouz2016discrete} &$O(\log k)$ &$O(\frac{\sqrt{k\log{k}}}{\epsilon\sqrt{N}})$   &$\frac{e^\epsilon +k-2}{N(e^\epsilon -1)^2}$  &N \\  \cline{2-6}
				&BLH \cite{wang2017protocol} &$O(\log k)$  &$O(\frac{\sqrt{\log{k}}}{\epsilon\sqrt{N}})$ &$\frac{(e^\epsilon+1)^2}{N(e^\epsilon-1)^2}$   &Y \\ \cline{2-6}
				&OLH \cite{wang2017protocol} &$O(\log k)$  &$O(\frac{\sqrt{\log{k}}}{\epsilon\sqrt{N}})$  &$\frac{4e^\epsilon}{N(e^\epsilon-1)^2}$  &Y \\ \hline
				\multirow{2}{*}{\textbf{\tabincell{l}{Transformation}}} &S-Hist \cite{bassily2015local} &$O(\log b)$ &$O(\frac{\sqrt{\log k}}{\epsilon\sqrt{N}})$  &$\frac{e^\epsilon}{N(e^\epsilon -1)^2}$  &Y \\ \cline{2-6}
				&HRR \cite{cormode2018marginal} &$O(\log k)$  &$O(\frac{\sqrt{\log{k}}}{\epsilon\sqrt{N}})$  &$\frac{4e^\epsilon}{N(e^\epsilon-1)^2}$ &Y \\ \hline
				\textbf{\tabincell{l}{Subset Selection}}&$\omega$-SM \cite{wang2016mutual, WangHNZWXY19} &$O(\omega)$ &$O(\frac{\sqrt{k\log{k}}}{\epsilon\sqrt{N}})$  &$\frac{e^\epsilon +k-2}{N(e^\epsilon -1)^2}$  &Y \\
				\toprule[1pt]
			\end{tabular}
		\end{threeparttable}
	}
\end{table*}

\textit{Discussions}. Section~\ref{sec-fre-basic} discusses the general frequency estimation protocols with LDP. When focusing on multiple attributes (i.e., $d$-dimensional data), we can directly use the above protocols to estimate the frequency of each attribute. We can also estimate joint frequency distributions of multiple attributes by using the above protocols as long as we regard the Cartesian product of the values of multiple attributes as the total domain. However, the error bound will be $d$ times greater than dealing with a single attribute. What's worse, the total domain cardinality will increase exponentially with the dimension $d$, which leads to huge computation overhead and low data utility. Therefore, several studies \cite{ren2016high, ren2018LoPub, yang2017copula, cormode2018marginal, zhang2018calm,wang2019locally} have investigated on estimating joint probability distributions of $d$-dimensional data, which are summarized in Section~\ref{sec-fre-marginal}.

\subsection{Frequency Estimation on Set-valued Data}\label{sec-fre-set-value}

This section summarizes the mechanisms for frequency estimation on set-valued data, including items distribution estimation, frequent items mining, and frequent itemsets mining.

\begin{wraptable}{r}{5cm}
\footnotesize
	\centering
	\caption{Example of \mbox{set-valued} dataset}\label{table-item}
	\setlength{\tabcolsep}{3mm}{
	\linespread{1}
	\begin{tabular}{c|c}
	\hline
	$V^1$ &$\{A,C,E\}$ \\\hline
	$V^2$ &$\{B,D,E\}$ \\\hline
	$V^3$ &$\{A,B,E\}$  \\\hline
	$V^4$ &$\{A,D,E\}$  \\\hline
	$V^5$ &$\{A,D,F\}$  \\\hline
	$V^6$ &$\{A,F\}$  \\\hline
	\end{tabular}}
\end{wraptable}

Set-valued data denotes a set of items. Let $\mathcal{K}=\{v_1,v_2,\cdots,v_k\}$ be the domain of items. Set-valued data $V^i$ of user $i$ is denoted as a subset of $\mathcal{K}$, i.e., $V^i \subseteq \mathcal{K}$. Different users may have different number of items. Table~\ref{table-item} shows a set-valued dataset of six users with item domain $\mathcal{K}=\{A,B,C,D,E,F\}$. In what follows, we will introduce each frequency-based task on set-valued data.

\subsubsection{Item Distribution Estimation}\label{sec-fre-set-value-fre}

The basic frequency estimation task on set-valued data is to analyze the distributions over $k$ items. For an item $v\in\mathcal{K}$, its frequency is defined as the fraction of times that $v$ occurs. That is, $f_v:=\frac{1}{N}|\{V^i|v\in V^i\}|$. Let $c_v$ be the number of users whose data include $v$. Then, we have $c_v:=|\{V^i|v\in V^i\}|$ and $f_v=\frac{1}{N}c_v$.

To tackle set-valued data with LDP, there are two tough challenges~\cite{wang2018privset}. (\textit{i}) \textit{Huge domain}: supposing there are total $k$ items and each user has at most $l$ items, then the number of possible combinations of items for each user could reach to $\binom{k}{l}+\binom{k}{l-1}+\cdots+\binom{k}{0}$. (\textit{ii}) \textit{Heterogeneous size}: different users may have different numbers of items, varying from $0$ to $l$. To address heterogeneous size, one of the most general methods is to add the padding items to the below-sized record. Then, the naive method is to treat the set-valued data as categorical data and use LDP algorithms in Section~\ref{sec-fre-basic}. However, this method needs to divide the privacy budget into smaller parts, thereby introducing excessive noise and reducing data utility.

Wang~\textit{et al.}~\cite{wang2018privset} proposed PrivSet mechanism which has a linear computational overhead with respect to the item domain size. PrivSet pre-processed the number of items of each user to $l$, which addresses the issue of heterogeneous size. When the item size is beyond $l$, it simply truncates or randomly samples $l$ items from the original set. When the item size is under $l$, it adds padding items to the original set. After pre-processing, each padded set-valued data $\widehat{s}$ belongs to $\widehat{S}=\{b|b\subseteq \widehat{\mathcal{K}} \text{~and~} |b|=l\}$, where $\widehat{\mathcal{K}}$ is the item domain after padding. Then, PrivSet randomized data based on the exponential mechanism. For each padded data $\widehat{s}\in \widehat{S}$, the randomization component of PrivSet selects and outputs an element $\widehat{t}\in \widehat{T}=\{a|a\subseteq \widehat{\mathcal{K}} \text{~and~} |a|=l'\}$ with probability
\begin{align}
p=
\begin{cases}
    \frac{exp(\epsilon)}{\Omega },&\text{~if~} \widehat{s}\cap\widehat{t}\neq\varnothing,\\
    \frac{1}{\Omega},&\text{~if~} \widehat{s}\cap\widehat{t}=\varnothing,
\end{cases}
\end{align}
where $\Omega$ is the probability normalizer and equals to $\binom{k}{l'}+exp(\epsilon)\cdot \left(\binom{k+l}{l'}-\binom{k}{l'}\right)$. As analyzed in \cite{wang2018privset}, the randomization of PrivSet reduces the computation cost from $O\left(\binom{k+l}{l}\right)$ to $O(k)$, which is linear to item domain size. PrivSet also holds a lower error bound over other mechanisms.

Moreover, LDPart \cite{zhao2019ldpart} is proposed to generate sanitized location-record data with LDP, where location-record data are treated as a special case of set-valued data. LDPart uses a partition tree to greatly reduce the domain space and leverages OUE \cite{wang2017protocol} to perturb the input values. However, the utility of LDPart quite relies on two parameters, i.e., the counting threshold and the maximum length of the record. It's very difficult to calculate the optimal values of these two parameters.

\subsubsection{Frequent Items Mining}\label{sec-fre-item}

Frequent items mining (also known as heavy hitters identification, top-$\omega$ hitters mining, or frequent terms discovery) has played important roles in data statistics. Based on the notations in Section~\ref{sec-fre-set-value-fre}, we say an item $v$ is $\omega$-heavy ($\omega$-frequent) if its multiplicity is at least $\omega$, i.e., $c_v\geq\omega$. The task of frequent items mining is to identify all $\omega$-heavy hitters from the collected data. For example, as shown in Table~\ref{table-item}, the $3$-heavy items are A, D, and E.

Bassily and Smith \cite{bassily2015local} focused on producing a succinct histogram that contains the most frequent items of the data under LDP. They leveraged the random matrix projection to achieve much lower communication cost and error bound than that of earlier methods in \cite{mishra2006privacy} and \cite{hsu2012distributed}. The specific process of S-Hist is introduced in Section \ref{sec-fre-basic}. Furthermore, a follow-up work in \cite{bassily2017practical} proposed TreeHist that computes heavy hitters from a large domain. TreeHist transforms the user's value into a binary string and constructs a binary prefix tree to compute frequent strings, which improves both efficiency and accuracy. Moreover, with strong theoretical analysis, Bun~\textit{et al.}~\cite{bun2018heavy} proposed a new heavy hitter mining algorithm that achieves the optimal worst-case error as a function of the domain size, the user number, the privacy budget, and the failure probability.

To address the challenge that the number of items in each user record is different, Qin~\textit{et al.}~\cite{qin2016heavy} proposed a Padding-and-Sampling frequency oracle (PSFO) that first pads user's items into a uniform length $l$ by adding some dummy items and then makes each user sample one item from possessed items with the same sampling rate. They designed LDPMiner based on RAPPOR \cite{erlingsson2014rappor} and S-Hist \cite{bassily2015local}. LDPMiner adopts a two-phase strategy with privacy budgets $\epsilon_1$ and $\epsilon_2$, respectively. In phase 1, LDPMiner identifies the potential candidate set of frequent items (i.e., the top-$\omega$ frequent items) by using a randomized protocol with $\epsilon_1$. The aggregator broadcasts the candidate set to all users. In phase 2, LDPMiner refines the frequent items from the candidates with the remaining privacy budget $\epsilon_2$ and outputs the frequencies of the final frequent items. LDPMiner is much wiser on budget allocation than naive method. However, LDPMiner still needs to split the privacy budget into $2l$ parts at both phases, which limits the data utility.

Wang~\textit{et al.}~ \cite{wang2019locally} proposed a prefix extending method (PEM) to discover heavy hitters from an extremely large domains (e.g., $k=2^{128}$). To address the computing challenge, PEM iteratively identifies the increasingly longer frequent prefixes based on a binary prefix tree. Specifically, PEM first divides users into $g$ equal-size groups, making each group $i(1\leq i \leq g)$ associate with a particular prefix length $s_i$, such that $1<s_1<s_i<\cdots<s_g= \left \lceil \log k \right \rceil $. Then each user reports the private value using LDP protocol and the server iterates through the groups. Obviously, group size $g$ is a key parameter that influences both the computation complexity and data utility. So, Wang~\textit{et al.}~ \cite{wang2019locally} further designed a sensitivity threshold principle that computes a threshold to control the false positives, thus maintaining the effectiveness and accuracy of PEM. Jia~\textit{et al.}~\cite{jia2019calibrate} have pointed out that prior knowledge can be used to improve the data utility of the LDP algorithms. Thus, \texttt{Calibrate} was designed to incorporate the prior knowledge via statistical inference, which can be appended to the existing LDP algorithms to reduce estimation errors and improve the data utility.

\subsubsection{Frequent Itemset Mining}\label{sec-fre-itemset}

Frequent itemset mining is much similar to frequent items mining, except that the desired results of the former will become a set of itemsets rather than items. Frequent itemsets mining is much more challenging since the domain size of itemsets is exponentially increased.

Following the definition of frequent item mining in Section~\ref{sec-fre-item}, the frequency of any itemset $\mathbf{v}\subseteq \mathcal{K}$ is defined as the fraction of times that itemset $\mathbf{v}$ occurs. That is, $f_{\mathbf{v}}:=\frac{1}{N}|\{V^i|\mathbf{v}\subseteq V^i\}|$. The count of any itemset $\mathbf{v}\subseteq \mathcal{K}$ is defined as the total number of users whose data include $\mathbf{v}$ as a subset. That is, $c_{\mathbf{v}}:=|\{V^i|\mathbf{v}\subseteq V^i\}|$. An $\omega$-heavy itemset $\mathbf{v}$ is such that its multiplicity is at least $\omega$, i.e., $c_{\mathbf{v}} \geq \omega$. The task of frequent itemset mining is to identify all $\omega$-heavy itemsets from the collected data. For example, the $3$-heavy itemsets in Table~\ref{table-item} are $\{A\}$, $\{D\}$, $\{E\}$, and $\{A,E\}$.

Sun~\textit{et al.}~\cite{sun2014personalized} proposed a personalized frequent itemset mining algorithm that provides different privacy levels for different items under LDP. They leveraged the randomized response technique \cite{warner1965randomized} to distort the original data with personalized privacy parameters and reconstructed itemset supports from the distorted data. This method distorted each item in domain separately, which leads to an error bound of $O(\frac{k\sqrt{\log k}}{\epsilon\sqrt{N}})$ that is super-linear to $k$. As introduced in Section~\ref{sec-fre-item}, LDPMiner mines frequent items over set-valued data by using a PSFO protocol. Inspired by LDPMiner, Wang~\textit{et al.}~\cite{wang2018locally} also padded users items into size $l$ and sampled one item from the possessed items of each user, which could ensure the frequent items can be reported with high probability even though there still exist unsampled items. Specifically, Wang~\textit{et al.}~\cite{wang2018locally} designed a Set-Value Item Mining (SVIM) protocol, and then based on the results from SVIM, they proposed a more efficient Set-Value ItemSet Mining (SVSM) protocol to find frequent itemsets. To further improve the data utility of SVSM, they also investigated the best-performing-based LDP protocol for each usage of PSFO by identifying the privacy amplification property of each LDP protocol.

\subsubsection{New Terms Discovery}\label{sec-fre-newterm}

This section introduces the task of new terms discovery that focuses on the situation where the global knowledge of item domain is unknown. Discovering top-$\omega$ frequent new terms is an important problem for updating a user-friendly mobile operating system by suggesting words based on a dictionary.

Apple iOS \cite{apple2017local,thakurta2017emoji}, macOS \cite{tang2017privacy}, and Google Chrome \cite{erlingsson2014rappor,fanti2016building} have integrated with LDP to protect users privacy when collecting and analyzing data. RAPPOR \cite{erlingsson2014rappor} is first used for frequency estimation under LDP, which is introduced in Section~\ref{sec-fre-basic}. Afterward, its augmented version A-RAPPOR \cite{fanti2016building} is proposed and applied in the Google Chrome browser for discovering frequent new terms. A-RAPPOR reduces the huge domain of possible new terms by collecting $n$-grams instead of full terms. Suppose the character domain is $C$. Then there will be $C/n$ such $n$-gram groups. For each group, A-RAPPOR constructs the significant $n$-grams that will be used to construct a $m$-partite graph, where $m=C/n$. Thus, the frequent new terms can be found efficiently by finding $m$-cliques in a $m$-partite graph. However, A-RAPPOR has rather low utility since the $n$-grams cannot always represent the real terms. The variance of each group is limited as $\frac{4e^\epsilon}{N(e^\epsilon -1)^2}$.
 
%Specifically, A-RAPPOR partitions the string into adjacent, non-overlapping $n$-grams.

To improve the accuracy and reduce the huge computational cost, Wang~\textit{et al.}~\cite{wang2018privtrie} proposed PrivTrie which leverages an LDP-complaint algorithm to iteratively construct a trie. When constructing a trie, the naive method is to uniformly allocate the privacy budget to each level of the trie. However, this will lead to inaccurate frequency estimations, especially when the height of the trie is large. To address this challenge, PrivTrie only requires to estimate a coarse-grained frequency for each prefix based on an adaptive user grouping strategy, thereby remaining more privacy budget for actual terms. PrivTrie further enforces consistency in the estimated values to refine the noisy estimations. Therefore, PrivTrie achieved much higher accuracy and outperformed the previous mechanisms. Besides, Kim \textit{et al.} \cite{kim2020learning} proposed a novel algorithm called CCE (Circular Chain Encoding) to discover new words from keystroke data under LDP. CCE leveraged the chain rule of $n$-grams and a fingerprint-based filtering process to improve computational efficiency.

\textit{Comparisons and discussions.} Table~\ref{compare-ldp-fre-set-value} shows the comparisons of LDP-based protocols for frequency estimation on set-valued data, including communication cost, key technique, and whether need to know the domain in advance. As for set-valued data, the padding-and-sampling technique is always adopted to solve the problem of the heterogeneous item size of different users, such as in \cite{wang2018privset,qin2016heavy,wang2018locally}. The padding size $l$ is a key parameter for both efficiency and accuracy. How to choose an optimal $l$ needs further study. Besides, tree-based method is also widely used in \cite{zhao2019ldpart,bassily2017practical,wang2019locally,wang2018privtrie} to reconstruct set-valued data. The tree-based method usually requires partition users into different groups or partition privacy budget for each level of the tree, which limits the data utility. In this case, the optimal budget allocation strategy needs to be designed. Meanwhile, the adaptive user grouping technique is also a better way to improve data utility, such as in \cite{wang2018privtrie}.

\begin{table*}[t]
\footnotesize
	\renewcommand\arraystretch{1} %行距
	\centering
	\caption{Comparisons of frequency estimation mechanisms for set-valued data with LDP}\label{compare-ldp-fre-set-value}
	\setlength{\tabcolsep}{1.5mm}{
		\fontsize{7pt}{\baselineskip}\selectfont
		\begin{threeparttable}
			\begin{tabular}{c|c|c|c|c}
				\bottomrule[1pt]
				\textbf{Task} &\textbf{LDP Algorithm} &\textbf{\tabincell{c}{Comm. Cost}} &\textbf{Key Technique} &\textbf{\tabincell{c}{Know\\Domain?}} \\ \hline%\xrowht{15pt}
				\multirow{2}{*}{\tabincell{c}{\textbf{Item distribution}\\\textbf{estimation}}} &PrivSet \cite{wang2018privset}  &$O(l')$~\tnote{1}  &Padding-and-sampling; Subset selection   &Y \\ \cline{2-5}%\xrowht{15pt}
				&LDPart \cite{zhao2019ldpart}  &$O(|V|_m)$~\tnote{2}    &Tree-based (partition tree); Users grouping  &Y \\ \hline%\xrowht{15pt}
				\multirow{4}{*}{\tabincell{c}{\textbf{Frequent}\\\textbf{items mining}}} &TreeHist \cite{bassily2017practical}  &$O(1)$  &Tree-based (binary prefix tree)  &Y  \\ \cline{2-5}%\xrowht{15pt}
				&LDPMiner \cite{qin2016heavy}  &$O(\log k +\omega)$ &Padding-and-sampling; Wiser budget allocation   &Y  \\ \cline{2-5}%\xrowht{15pt}
				&PEM \cite{wang2019locally}  &$O(\log k)$  &Tree-based (binary prefix tree); Users grouping   &Y \\ \cline{2-5}%\xrowht{15pt} 
				&\texttt{Calibrate} \cite{jia2019calibrate}  &$O(k)$ &Consider prior knowledge   &Y  \\ \hline%\xrowht{15pt}
				\multirow{2}{*}{\tabincell{c}{\textbf{Frequent}\\\textbf{itemset mining}}} &Personalized \cite{sun2014personalized} &$O(k)$  &Personalized privacy regime  &Y  \\ \cline{2-5}%\xrowht{15pt}
				&SVSM \cite{wang2018locally}  &$O(\log k)$  &Padding-and-sampling; Privacy amplification  &Y \\ \hline%\xrowht{15pt}
				\multirow{3}{*}{\tabincell{c}{\textbf{New terms}\\\textbf{discovering}}}  &A-RAPPOR \cite{fanti2016building}  &$O(\log k)$  &Select $n$-grams; Construct partite graph  &N \\ \cline{2-5}%\xrowht{15pt}
				&PrivTrie \cite{wang2018privtrie}  &$O(|V|_m)$  &\tabincell{c}{Tree-based (trie); Adaptive users grouping;\\Consistency constraints}  &N \\
				%&CCE \cite{kim2020learning} &   &    &N \\
				\toprule[1pt]
			\end{tabular}
			\begin{tablenotes}
				\footnotesize
				\item[1] $l'$ is the output size of randomization, which is smaller than total domain size $k$.
				\item[2] $|V|_m$ is the maximum number of nodes among all layers of the tree.
			\end{tablenotes}
		\end{threeparttable}
	}
\end{table*}

\subsection{Frequency Estimation on Key-Value Data}\label{sec-fre-key-value}

Key-value data~\cite{ye2019privkv} is such data that has a key-value pair including a key and a value, which is commonly used in big data analysis. For example, the key-value pairs (KV pair) that denote diseases and their diagnosis values are listed as $\{\left \langle Cancer,0.3 \right \rangle\}$, $\{\left \langle Fever,0.06 \right \rangle\}$, $\{\left \langle Flu,0.6 \right \rangle\}$, etc. While collecting and analyzing key-value data, there are four challenges to consider. (\textit{i}) Key-value data contain two heterogeneous dimensions. The existing studies mostly focus on homogeneous data. (\textit{ii}) There are inherent correlations between keys and values. The naive method that deals with key-value data by separately estimating the frequency of key and the mean of value under LDP will lead to the poor utility. (\textit{iii}) One user may possess multiple key-value pairs that need to consume more privacy budget, resulting in larger noise. (\textit{iv}) The overall correlated perturbation mechanism on key-value data should consume less privacy budget than two independent perturbation mechanisms for key and value respectively. We can improve data utility by computing the actually consumed privacy budget.

Ye \textit{et al.}~\cite{ye2019privkv} proposed $PrivKV$ that retains the correlations between keys and values while achieving LDP. $PrivKV$ adopts $Harmony$ \cite{nguyen2016collecting} to perturb the value $v$ of a KV pair into $v^*$ with privacy budget $\epsilon_2$ and converts the pair $\left \langle key, v^* \right \rangle$ into canonical form $\left \langle 1, v^* \right \rangle$ that is perturbed by
\begin{align}
    \left \langle key, v^* \right \rangle=
    \begin{cases}
        \left \langle 1, v^* \right \rangle, &\text{w.p.~}\frac{e^{\epsilon_1}}{e^{\epsilon_1}+1},\\
        \left \langle 0, 0 \right \rangle, &\text{w.p.~}\frac{1}{e^{\epsilon_1}+1}.
    \end{cases}
\end{align}

Noted that in $PrivKV$, the value of a key-value pair is randomly drawn from the domain of $[-1, 1]$ when users don't own a key-value pair (i.e., the users have no apriori knowledge about the distribution of true values). Thus, PrivKV suffers from low accuracy and instability. Therefore, Ye~\textit{et al.}~\cite{ye2019privkv} built two algorithms $PrivKVM$ and $PrivKVM^+$ by multiple iterations to address this problem. Intuitively, as the number of iterations increases, the accuracy will be improved since the distribution of the values will be close to the distributions of the true values.

Based on $PrivKV$, Sun~\textit{et al.}~\cite{sun2019conditional} proposed several mechanisms for key-value data collection based on direct encoding and unary encoding techniques \cite{wang2017protocol}. Sun~\textit{et al.}~\cite{sun2019conditional} also introduced conditional analysis for key-value data for the first time. Specifically, they proposed several mechanisms that support $L$-way conditional frequency and mean estimation while ensuring good accuracy.

However, the studies in \cite{ye2019privkv} and \cite{sun2019conditional} lack exact considerations on challenges (\textit{iii}) and (\textit{iv}) mentioned previously. On the one hand, they simply sample a pair when facing multiple key-value data pairs, which cannot make full use of the whole data pairs and may not work well for a large domain. On the other hand, they neglect the privacy budget composition when considering the inherent correlations of key-value data, thus leading to limited data utility.

Gu~\textit{et al.}~\cite{gu2019pckv} proposed the correlated key/value perturbation mechanism which reduces the privacy budget consumption and enhances data utility. They designed a Padding-and-Sampling protocol for key-value data to deal with multiple pairs of each user. Thus, it is no longer necessary to sample a pair from the whole domain (e.g., PrivKVM\cite{ye2019privkv}), but sample from the key-value pairs possessed by users, thus eliminating the affects of large domain size. Then, they proposed two protocols PCKV-UE by adopting unary encoding and PCKV-GRR by adopting the generalized randomized response. Rather than sequential composition, both PCKV-UE and PCKV-GRR involve a near-optimal correlated budget composition strategy, thereby minimizing the combined mean square error.

\begin{table*}[tb]
	\centering
   \footnotesize
	\caption{Comparisons of LDP-based protocols for frequency/mean estimation on key-value data}\label{compare-ldp-fre-key-value}
	\setlength{\tabcolsep}{1.5mm}{
		\begin{threeparttable}
			\begin{tabular}{c|c|c|c|c|c}
				\bottomrule[1pt]
				\textbf{LDP Algorithm} &\textbf{Goal}  &\textbf{Address Multiple Pairs} &\tabincell{c}{\textbf{Learn}\\\textbf{Correlations}} &\textbf{Composition} &\textbf{Allocation of $\epsilon$} \\ \hline
				$PrivKVM$~\cite{ye2019privkv} &\tabincell{c}{Mean value of values;\\frequency of keys} &Simple sampling &\tabincell{c}{Mechanism\\iteration} &Sequential &Fixed  \\\hline
				CondiFre~\cite{sun2019conditional} &\tabincell{c}{Mean value of values;\\frequency of keys;\\$L$-way conditional analysis} &Simple sampling &Not consider &Sequential &Fixed  \\\hline
				\tabincell{c}{PCKV-UE/\\PCKV-GRR~\cite{gu2019pckv} }&\tabincell{c}{Mean value of values;\\frequency of keys} &Padding-and-sampling &\tabincell{c}{Correlated\\perturbation} &Tighter bound &Optimal  \\
				\toprule[1pt]
			\end{tabular}
		\end{threeparttable}
	}
\end{table*}

\textit{Comparisons and discussions.} Table~\ref{compare-ldp-fre-key-value} shows the comparisons of frequency/mean estimations on key-value data with LDP. To solve the challenge of each user has multiple data pairs, simple sampling \cite{ye2019privkv,sun2019conditional} is adopted to sample a pair from the whole domain and padding-and sampling \cite{gu2019pckv} is adopted to sample a pair from the key-value pairs possessed by users. Besides, $PrivKVM$~\cite{ye2019privkv} and PCKV-UE/PCKV-GRR~\cite{gu2019pckv} consider the correlations between key and value by iteration and correlated perturbation, respectively. But CondiFre~\cite{sun2019conditional} lacks of the consideration on learning correlations when conducting conditional analysis. What's more, both $PrivKV$~\cite{ye2019privkv} and CondiFre~\cite{sun2019conditional} achieve LDP based on two independent perturbations with fixed privacy budget by sequential composition. PCKV-UE/PCKV-GRR~\cite{gu2019pckv} holds a tighter privacy budget composition strategy that makes the optimal allocation of privacy budget.

\subsection{Frequency Estimation on Ordinal Data}\label{sec-fre-ordinal}

Compared to categorical data, ordinal data has a linear ordering among categories, which is concluded as ordered categorical data, discrete numerical data (e.g., discrete sensor/metering data), and preference ranking data makes the optimal allocation of privacy budget. 

When quantifying the indistinguishability of two ordinal data with LDP, the work in \cite{andres2013geo} measured the distance of two ordinal data by $\epsilon$-geo-indistinguishability. That is, a mechanism satisfies $\epsilon$-DP if it holds $\mathbb{P}[\mathcal{M}(X_i) \in \mathcal{Y}] \leq e^{\epsilon\cdot d(X_i,X_j)} \cdot \mathbb{P}[\mathcal{M}(X_j) \in \mathcal{Y}]$ for any possible pairs $X_i,X_j\in \mathcal{X}$. The distance $d(X_i,X_j)$ of ordinal data $X_i$ and $X_j$ can be measured by Manhattan distance or squared Euclidean distance. Based on $\epsilon$-geo-indistinguishability, Wang~\textit{et al.}~\cite{wang2017local,WangHNZWXY19} proposed subset exponential mechanism (SEM) that is realized by a tweaked version of exponential mechanism \cite{dwork2014algorithmic}. Besides, they also proposed a circling subset exponential mechanism (CSEM) for ordinal data with uniform topology. For both SEM and CSEM, the authors have provided the theoretical error bounds to show their mechanisms can reduce nearly a fraction of $\exp(-\frac{\epsilon}{2})$ error for frequency estimation.

Preference ranking data is also one of the most common representations of personal data and highly sensitive in some applications, such as preference rankings on political or service quality. Essentially, preference ranking data can be regarded as categorical data that hold an order among different items. Given an item set $\mathcal{K}=\{v_1,v_2,\cdots,v_k\}$, a preference ranking of $\mathcal{K}$ is an ordered list that contains all $k$ items in $\mathcal{K}$. Denote a preference ranking as $\sigma =\left \langle \sigma (1), \sigma (2), \cdots,\sigma (k), \right \rangle$, where $\sigma (j)=v_i$ means that item $v_i$'s rank under $\sigma$ is $j$. The goal of collecting preference rankings is to estimate the distribution of all different rankings from $N$ users. It can easily verify that the domain of all rankings is $k!$, which leads to excessive noises and low accuracy when $k$ is large. Yang~\textit{et al.}~\cite{yang2019collecting} proposed SAFARI that approximates the overall distributions over a smaller domain that is chosen based on a riffle independent model. SAFARI greatly reduces the noise amount and improves data utility.

Voting data is to some extent a kind of preference ranking. By aggregating the preference rankings based on one of the certain positional voting rules (e.g., Borda, Nauru, Plurality \cite{black2012theory,reilly2002social,brandt2016handbook}), we can obtain the collective decision makings. To avoid leaking personal preferences in a voting system, the work in \cite{wang2019aggregating} collected and aggregated voting data with LDP while ensuring the usefulness and soundness. Specifically, weighted sampling mechanism and additive mechanism are proposed for LDP-based voting aggregation under general positional voting rules. Compared to the na\"ive Laplace mechanism, weighted sampling mechanism and additive mechanism can reduce the maximum magnitude risk bound from $+\infty$ to $O(\frac{k^3}{N\epsilon})$ and $O(\frac{k^2}{N\epsilon})$, respectively, where $k$ is the size of vote candidates (i.e., the domain size), $N$ is the number of users.

\iffalse

\begin{table*}[t]
\centering
\caption{Comparisons of LDP-based protocols for frequency estimation on ordinal data}\label{compare-ldp-fre-ordinal}
\setlength{\tabcolsep}{1.5mm}{
\fontsize{8pt}{\baselineskip}\selectfont
\begin{threeparttable}
\begin{tabular}{c|c|c|c|c}
\bottomrule[1pt]
\textbf{LDP Algorithms} &\textbf{Communication Cost} &\textbf{Error Bound} &\textbf{Variance} &\textbf{\tabincell{c}{Know\\Domain?}} \\ \hline
SEM \cite{wang2017local,WangHNZWXY19} & &  & & \\ \hline
CSEM \cite{wang2017local,WangHNZWXY19} & & &  & \\ \hline
SAFARI \cite{yang2019collecting}  & & & &  \\ \hline
Weighted Sampling \cite{wang2019aggregating} & & & &  \\ \hline
Additive \cite{wang2019aggregating}  & & & &  \\
\toprule[1pt]
\end{tabular}
\end{threeparttable}
}
\end{table*}

\fi
%point query

As one of the fundamental data analysis primitives, range query aims at estimating the fractions or quantiles of the data within a specified interval \cite{li2014data,alnemari2017adaptive}, which is also an analysis task on ordinal data. The studies in \cite{tejas2019answering,cormode2019answering} have proposed some approaches to support range queries with LDP while ensuring good accuracy.  They designed two methods to describe and analyze the range queries based on hierarchical histograms and the Haar wavelet transform, respectively. Both methods use OLH \cite{wang2017protocol} to achieve LDP with low communication cost and high accuracy. Besides, local \texttt{d}-privacy is a generalized notion of LDP under distance metric, which is adopted to assign different perturbation probabilities for different inputs based on the distance metrics. Gu~\textit{et al.}~\cite{gu2019supporting} used local \texttt{d}-privacy to support both range queries and frequency estimation. They proposed an optimization framework by solving the linear equations of perturbation probabilities rather than solving an optimization problem directly, which not only reduces computation cost but also makes the optimization problem always solvable when using \texttt{d}-privacy.

\subsection{Frequency Estimation on Numeric Data}\label{sec-fre-numeric-distribution}

Most existing studies compute frequency estimations on categorical data. However, there are many numerical attributes in nature, such as income, age. Computing frequency estimation of numeric data also plays important role in reality. 

For numerical distribution estimation with LDP, the na\"ive method is to discretize the numerical domain and apply the general LDP protocols directly. However, the data utility of the na\"ive method relies heavily on the granularity of discretization. What's worse, an optimal discretization strategy depends on privacy parameters and the original distributions of the numeric attributes. Thus, it's a big challenge to find the optimal discretization strategy. Li~\textit{et al.}~\cite{li2020estimating} utilized the ordered nature of the numerical domain to compromise a better trade-off between privacy and utility. They proposed a novel mechanism based on expectation-maximization and smoothing techniques, which improves the data utility significantly.

\subsection{Marginal Release on Multi-dimensional Data}\label{sec-fre-marginal}
The marginal table is ``the workhorse of data analysis'' \cite{cormode2018marginal}. When obtaining marginal tables of a set of attributes, we can learn the underlying distributions of multiple attributes, identify the correlated attributes, describe the probabilistic relationships between cause and effects. Thus, $k$-way marginal release has been widely investigated with LDP \cite{yang2017copula,ren2018LoPub,cormode2018marginal}.

Denote $A=\{A_1,A_2,\cdots,A_d\}$ as the $d$ attributes of $d$-dimensional data. For each attribute $A_j$ $(j=1,2,\cdots,d)$, the domain of $A_j$ is denoted as $\Omega_j=\{\omega_j^1, \omega_j^2, \cdots, \omega_j^{|\Omega_j|}\}$, where $\omega_j^i$ is the $i$-th value of $\Omega_j$ and $|\Omega_j|$ is the cardinality of $\Omega_j$. The marginal table is defined as follows.

\begin{defn}[Marginal Table \cite{cormode2018marginal}] \label{defn-mar-tab}
Given $d$-dimensional data, marginal operator $\mathcal{C}^\beta$ computes all frequencies of different attribute combinations that are decided by $\beta \in \{0,1\}^d$, where $\left|\beta\right|$ denotes the number of $1'$s in $\beta$, and $\left|\beta\right|=k\leq d$. The marginal table contains all the returned results of $\mathcal{C}^\beta$.
\end{defn}

\textit{Example 1.} When $d=4$ and $\beta=0110$, it means that we estimate the probability distributions of all combination of the second and third attributes. $\mathcal{C}^{0110}$ returns the frequency distributions of all combinations.

The $k$-way marginal is the probability distributions of any $k$ attributes in $d$ attributes. 

\begin{defn}[$k$-way Marginal \cite{cormode2018marginal}]
The \textit{k-way} marginal is the probability distributions of $k$ attributes in $d$ attributes, that is, $|\beta|=k$. For a fixed $k$, the set of all possible $k$-way marginals correspond to all $\binom{d}{k}$ distinct ways of picking $k$ attributes from $d$, which called full $k$-way marginals. 
\end{defn}

The $k$-way marginal release is to estimate $k$-way marginal probability distribution for any $k$ attributes $A_{j_1}, A_{j_2}, \cdots, A_{j_k}$ chosen from $A$. The $k$-way marginal distribution of attributes $A_{j_1}, A_{j_2}, \cdots, A_{j_k}$ is denoted as $P(A_{j_1}A_{j_2}\cdots A_{j_k})$. It has
\begin{align}
    P(A_{j_1}A_{j_2}\cdots A_{j_k})\triangleq P(\omega_{j_1}\omega_{j_2}\cdots\omega_{j_k})\\\text{for~}\forall\omega_{j_1}\in\Omega_{j_1},\omega_{j_2}\in\Omega_{j_2},\cdots,\omega_{j_k}\in\Omega_{j_k}.\nonumber
\end{align}

\subsubsection{$k$-way Marginal Probability Distribution Estimation}\label{sec-fre-k-way}

Randomized response technique \cite{erlingsson2014rappor} can be na\"ively leveraged to achieve LDP when computing $k$-way marginal probability distributions. However, both the efficiency and accuracy will be seriously affected by the ``curse of high-dimensionality''. The total domain cardinality will be $\prod _{j=1}^{k}|\Omega_j|$, which increases exponentially as $k$ increases.

The EM-based algorithm with LDP \cite{fanti2016building} is restricted to 2-way marginals. When $k$ is large, it will lead to high time/space overheads. The work in~\cite{ren2016high} proposed a Lasso-based regression mechanism that can estimate high-dimensional marginals efficiently by extracting key features with high probabilities. Besides, Ren~\textit{et al.}~\cite{ren2018LoPub} proposed LoPub to find compactly correlated attributes to achieve dimensionality reduction, which further reduces the time overhead and improves data utility. 

Nonetheless, the $k$-way marginals release still suffers from low data utility and high computational overhead when $k$ becomes larger. To solve this, the work in~\cite{yang2017copula} proposed to leverage Copula theory to synthesize multi-dimensional data with respect to marginal distributions and attribute dependence structure. It only needs to estimate one and two-marginal distributions instead of $k$-way marginals, thus circumventing the exponential growth of domain cardinality and avoiding the curse of dimensionality. Afterward, Wang~\textit{et al.}~\cite{wang2019locallycopula} further leveraged C-vine Copula to take the conditional dependencies among high-dimensional attributes into account, which significantly improves data utility.

Cormode~\textit{et al.}~\cite{cormode2018marginal} have investigated marginal release under under different kinds of LDP protocols. They further proposed to materialize marginals by a collection of coefficients based on the Hadamard transform (HT) technique. The underlying motivation of using HT is that the computation of $k$-way marginals require only a few coefficients in the Fourier domain. Thus, this method improves the accuracy and reduces the communication cost. Nonetheless, this method is designed for binary attributes. The non-binary attributes need to be pre-processed to binary types, leading to higher dimensions. To further improve the accuracy, Zhang~\textit{et al.}~\cite{zhang2018calm} proposed a consistent adaptive local marginal (CALM) algorithm. CALM is inspired by PriView \cite{qardaji2014priview} that builds $k$-way marginal by taking the form of $m$ marginals each of the size $l$ (i.e., synopsis). Besides, the work in \cite{wang2019answering, xu2019dpsaas} focused on answering multi-dimensional analytical queries that are essentially formalized as computing $k$-way marginals with LDP.

\textit{Comparisons and discussions}. Table~\ref{compare-ldp-fre-marginal} summarized the LDP-based algorithms for $k$-way marginal release. To improve efficiency, the existing methods try to reduce the large domain space by various techniques, such as HT and dimensionality reduction. As we can see, the variances of the existing methods are relatively large, leading to limited data utility. Although the subset selection is a useful way to reduce the communication cost and variance, it suffers from the sampling error when constructing low-dimensional synopsis. Therefore, designing mechanisms with high data utility and low costs still faces big challenges when $d$ is large.

\begin{table*}[tb]
\footnotesize
	\centering
	\caption{Comparisons of LDP-based algorithms for $k$-way marginal release of $d$-dimensional data}\label{compare-ldp-fre-marginal}
	\setlength{\tabcolsep}{1mm}{
		\begin{threeparttable}
			\begin{tabular}{c|c|c|c|c|c}
				\bottomrule[1pt]
				\textbf{LDP Algorithm} &\textbf{Key Technique}  &\textbf{Comm. Cost} &\textbf{Variance}  &\textbf{Time Complexity} &\textbf{Space Complexity} \\ \hline
				RAPPOR \cite{erlingsson2014rappor} &\tabincell{c}{Equal to na\"ive\\method when $d$>2 } &$O\left(\prod_{j=1}^d|\Omega_j| \right)$ &$2^d \cdot$Var \tnote{1} &High &High\\\hline
				Fanti \textit{et al.}~\cite{fanti2016building} &\tabincell{c}{Expectation\\Maximization (EM)} &$O\left(\sum_{j=1}^d|\Omega_j| \right)$ &$2^d \cdot$Var &$O\left(N\cdot\sum_{i=1}^{k}\binom{d}{i}2^i\right)$ &$O\left(\sum_{i=1}^{k}\binom{d}{i}2^i\right)$  \\\hline
				LoPub~\cite{ren2018LoPub} &\tabincell{c}{Lasso regression; Dimensionality\\and sparsity reduction} &$O\left(\sum_{j=1}^d|\Omega_j| \right)$ &$2^d \cdot$Var &Medium &High  \\\hline
				Cormode \textit{et al.} \cite{cormode2018marginal}  &\tabincell{c}{Hadamard\\Transformation (HT)} &$O\left( \sum_{i=1}^k \binom{d}{i} \right)$ &$\sum_{i=1}^k \binom{d}{i}\cdot$Var &$O\left(N+\binom{d}{k}\cdot2^k\right)$ &$O\left(\sum_{i=1}^{k}\binom{d}{i}\right)$\\\hline
				LoCop~\cite{wang2019locallycopula} &\tabincell{c}{Lasso-based regression;\\Attribute correlations learning} &$O\left(\sum_{j=1}^d|\Omega_j| \right)$ &$2^d \cdot$Var &Low &High  \\\hline
				CALM \cite{zhang2018calm} &\tabincell{c}{Subset selection;\\Consistency constraints}  &$O(2^l)$~\tnote{2} &$\frac{m}{N}\cdot 2^l \cdot$Var &$O(N\cdot 2^l)$ &$O(m\cdot 2^l)$\\
				\toprule[1pt]
			\end{tabular}
			\begin{tablenotes}
				\footnotesize
				\item[1] Var is the variance of estimating a single cell in the full contingency table;\vspace{2mm}
				\item[2] $l$ is the size of $m$ low marginals of dataset.
			\end{tablenotes}
		\end{threeparttable}
	}
\end{table*}

\subsubsection{Conditional Probability Distribution Estimation}\label{sec-fre-condi}

The conditional probability is also important for statistics. Sun~\textit{et al.}~\cite{sun2019conditional} have investigated on conditional distribution estimation for the keys in key-value data. They formalized $k$-way conditional frequency estimation and applied the advanced LDP protocols to compute $k$-way conditional distributions. Besides, Xue \textit{et al.} \cite{xue2019joint} proposed to compute the conditional probability based on $k$-way marginals and further train a Bayes classifier.

\subsection{Frequency Estimation on Evolving Data}\label{sec-fre-envolving}

So far, most academic literature focuses on frequency estimation for one-time computation with LDP. However, privacy leaks will gradually accumulate as time continues to grow under centralized DP \cite{cao2017quantifying,wang2017cts,cao2019quantifying}, so does LDP \cite{ding2017collecting,joseph2018local}. Therefore, when applying LDP for dynamic statistics over time, an LDP-compliant method should take time factor into account. Otherwise, the mechanism is actually difficult to achieve the expected privacy protection over long time scales. For example, Tang \textit{et al.} \cite{tang2017privacy} have pointed out that the privacy parameters provided by Apple's implementation on MacOS will actually become unreasonably large even in relatively short time periods. Therefore, it requires careful considerations for longitudinal attacks on evolving data.

Erlingsson \textit{et al.} \cite{erlingsson2014rappor} adopted a heuristic memoization technique to provide longitudinal privacy protection in the case that multiple records are collected from the same user. Their method includes Permanent randomized response and Instantaneous randomized response. These two procedures will be performed in sequence with a memoization step in between. The Permanent randomized response outputs a perturbed answer which is reused as the real answer. The Instantaneous randomized response reports the perturbed answer over time, which prevents possible tracking externalities. In particular, the longitudinal privacy protection in work \cite{erlingsson2014rappor} assumes that the user value does not change over time, like the continual observation model \cite{dwork2010differential}. Thus, the approach in \cite{erlingsson2014rappor} cannot guarantee strong privacy for the users who have numeric values with frequent changes.

Inspired by \cite{erlingsson2014rappor}, Ding \textit{et al.} \cite{ding2017collecting} used the permanent memoization for continual counter data collection and histogram estimation. They first designed $\omega$-bit mechanism $\omega$\textsf{BitFlip} to estimate frequency of counter values in a discrete domain with $k$ buckets. In $\omega$\textsf{BitFlip}, each user randomly draws $\omega$ bucket numbers without replacement from $[k]$, denoted as $j_1,j_2,\cdots,j_\omega$. At each time $t$, each user randomizes her data $v_t\in [k]$ and reports a vector $b_t=[(j_1, b_t(j_1)), (j_2, b_t(j_2)), \cdots, (j_\omega, b_t(j_\omega))]$, where $b_t(j_z)$ ($z=1,2,\cdots,\omega$) is a random 0/1 bit with
\begin{align}
	\mathbb{P}[b_t(j_z) = 1] = 
	\begin{cases}
	\frac{e^{\epsilon/2}}{e^{\epsilon/2} + 1}, \text{~if~}v_t = j_z, \\
	\frac{1}{e^{\epsilon/2} + 1}, \text{~if~}v_t \neq j_z.
	\end{cases}
\end{align}
Assume the sum of received 1 bit is $\bar{N}_{v_t}$. Then, the estimated frequency of $v_t\in [k]$ is
\begin{align}
	\widehat{f_{v_t}} = \frac{k}{N\omega}\sum_{\bar{N}_{v_t}}\frac{\bar{N}_{v_t} \cdot (e^{\epsilon/2}) -1}{e^{\epsilon/2} - 1}.
\end{align}

As we can see, $\omega$\textsf{BitFlip} will be the same as the one in Duchi \textit{et al.} \cite{duchi2018minimax} when $\omega = k$. Ding \textit{et al.} \cite{ding2017collecting} have proved that $\omega$\textsf{BitFlip} holds an error bound of $O\left(\frac{\sqrt{k\log k}}{\epsilon \sqrt{N\omega}}\right)$. 

In na\"ive memoization, each user reports a perturbed value based on the mapping $f_k:[k] \rightarrow \{0,1\}^k$, which leads to privacy leakage. To tackle this, Ding \textit{et al.} \cite{ding2017collecting} proposed $\omega$-bit permanent memoization mechanism $\omega$\textsf{BitFlipPM} based on $\omega$\textsf{BitFlip}. $\omega$\textsf{BitFlipPM} reports each response in a mapping $f_\omega:[k] \rightarrow \{0,1\}^\omega$, which avoids the privacy leakage since multiple buckets are mapped to the same response. 

Moreover, Joseph \textit{et al.} \cite{joseph2018local} proposed a novel LDP-compliant mechanism \textsc{THRESH} for collecting up-to-date statistics over time. The key idea of \textsc{THRESH} is to update the global estimation only when it might become sufficiently inaccurate. To identify these update-needed epochs, Joseph \textit{et al.} designed a \textit{voting protocol} that requires users to privately report a vote for whether they believe the global estimation needs to be updated. The \textsc{THRESH} mechanism can ensure that the privacy guarantees only degrade with the number of times of the statistics changes, rather than the number of times the computation of the statistics. Therefore, it can achieve strong privacy protection for frequency estimation over time while ensuring good accuracy.

\section{Mean Value Estimation with LDP}\label{sec-mean}
This Section summarizes the task of mean value estimation for numeric data with LDP, including mean value estimation on numeric data and mean value estimation on evolving data.

In formal, let $\mathcal{D}=\{V^1,V^2,\cdots,V^N\}$ be the data of all users, where $N$ is the number of users. Each tuple $V^i= (v_1^i,v_2^i,\cdots,v_d^i )$ $(i\in[1,N])$ denotes the data of the $i$-th user, which consists of $d$ numeric attributes $A_1,A_2,\cdots,A_d$. Each $v_j^i$ $(j\in[1,d])$ denotes the value of the $j$-th attribute of the $i$-th user. Without loss of generality, the domain of each numeric attribute is normalized into $[-1,1]$. The mean estimation is to estimate the mean value of each attribute $A_j(j\in[1,d])$ over $N$ users, that is $\frac{1}{N}\sum_{i=1}^{N}v_j^i$.

\subsection{Mean Value Estimation on Numeric Data}\label{sec-mean-numeric}

Let $\widehat{V}^i=(\widehat{v}_1^i,\widehat{v}_2^i,\cdots,\widehat{v}_d^i)$ be the perturbed $d$-dimensional data of user $i$. Given a perturbation mechanism $\mathcal{M}$, we use $\mathbb{E}[\widehat{v}_j]$ to denote the expectation of the output $\widehat{v}_j$ given an input $v_j$. Therefore, to achieve LDP, a perturbation mechanism should satisfy the following two constraints, that is,
\begin{align}
	&\mathbb{E}[\widehat{v}_j]=v_j,\label{constaint1} \\
	&\mathbb{P}[\widehat{v}_j \in \mathbb{V}|v] = 1.\label{constaint2} 
\end{align}
The first constraint (i.e., Eq. (\ref{constaint1})) shows that the mechanism should be unbiased. The second constraint (i.e., Eq. (\ref{constaint2})) shows that the sum of probabilities of the outputs must be one, where $\mathbb{V}$ is the output range of $\mathcal{M}$.

Laplace mechanism \cite{dwork06Calibrating} under DP can be applied in a distributed manner to achieve LDP. Based on Laplace mechanism, each user's data will be perturbed by adding randomized Laplace noise, that is, $\widehat{V}^i=V^i + \langle Lap(\frac{d\Delta}{\epsilon})\rangle^d$, where $Lap(\lambda)$ is a random variable drawn from a Laplace distribution with the probability density function of $pdf(v)=\frac{1}{2\lambda}\exp{-\frac{|v|}{\lambda}}$. Note that $\Delta$ is the sensitivity and $\Delta=2$ since each numeric data lies in range $[-1,1]$. The privacy budget for each dimension is $\epsilon/d$. Then, the aggregator will compute the average value of all received noisy reports as $\frac{1}{N}\sum_{i=1}^N\widehat{v}_j^i$. It can be easily verified that $\frac{1}{N}\sum_{i=1}^N\widehat{v}_j^i$ is an unbiased estimator of the $j$-th attribute since the injected Laplace noises have zero mean. Thus, the final mean value is also unbiased. Besides, the variance of the estimated $\widehat{v}_i$ is $\frac{8d^2}{\epsilon^2}$. The amount of noise of the estimated mean of each attribute is $O(\frac{d\sqrt{\log d}}{\epsilon\sqrt{N}})$, which is super-linear to dimension $d$. When $d=1$, the variance is $\frac{8}{\epsilon^2}$ and the error bound is $O(\frac{1}{\epsilon\sqrt{N}})$. As we can see, the Laplace mechanism will incur excessive error when $d$ becomes large.

Duchi \textit{et al.} \cite{duchi2018minimax} proposed an LDP-compliant method for collecting multi-dimensional numeric data. The basic idea of Duchi \textit{et al.}'s method is to utilize a randomized response technique to perturb each user's data according to a certain probability distribution while ensuring an unbiased estimation. Each user's tuple $V^i\in [-1,1]^d$ will be perturbed into a noisy vector $\widehat{V}^i\in \{-B, B\}^d$, where $B$ is a constant decided by $d$ and $\epsilon$. According to \cite{duchi2018minimax}, $B$ is computed as
\begin{align}
    B=
    \begin{cases}
    \frac{2^d+C_d\cdot(e^\epsilon-1)}{\binom{d-1}{(d-1)/2}\cdot(e^\epsilon -1)},&\text{~if~}d\text{~is odd},\vspace{1mm}\\
    \frac{2^d+C_d\cdot(e^\epsilon-1)}{\binom{d-1}{d/2}\cdot(e^\epsilon -1)},&\text{~otherwise},
    \end{cases}
\end{align}
where
\begin{align}
    C_d=
    \begin{cases}
    2^{d-1},&\text{~if~}d\text{~is odd},\\
    2^{d-1}-\frac{1}{2}\binom{d}{d/2},&\text{~otherwise}.
    \end{cases}
\end{align}

Duchi \textit{et al.}'s method takes a tuple $V^i\in [-1,1]^d$ as inputs and discretizes the $d$-dimensional data into $X : = [X_1, X_2, \cdots, X_d] \in \{-1,1\}^d$ by sampling each $X_j$ independently from the following distribution.
\begin{align}
    \mathbb{P}[X_j=x_j]=\begin{cases}
    \frac{1}{2}+\frac{1}{2}v_j^i,~~\text{if}~~x_j=1,\\
    \frac{1}{2}-\frac{1}{2}v_j^i,~~\text{if}~~x_j=-1.
    \end{cases}
\end{align}
In the case of $X$ is sampled, let $T^+$ (resp. $T^-$) be the set of all tuples $\widehat{V}^i\in\{-B,B\}^d$ such that $\widehat{V}^i\cdot X > 0$ (resp. $\widehat{V}^i\cdot X\leq 0$). The algorithm will return a noisy value based on the value of a Bernoulli variable $u$. That is, it will return $\widehat{V}^i$ uniformly at random from $T^+$ with probability of $\bp{u=1}=\frac{e^\epsilon}{e^\epsilon +1}$ or return a noisy value $\widehat{V}^i$ uniformly at random from $T^-$ with probability of $\bp{u=0}=\frac{1}{e^\epsilon +1}$.

Duchi \textit{et al.} have shown that $\frac{1}{N}\sum_{i=1}^N\widehat{v}_j^i$ is an unbiased estimator for each attribute $A_j$. Besides, the error bound of Duchi \textit{et al.}'s method is $O\left(\frac{\sqrt{d\log d}}{\epsilon\sqrt{N}}\right)$.

Although Duchi \textit{et al.}'s method can achieve LDP and has an asymptotic error bound, it's relatively sophisticated. Nguy{\^e}n~\textit{et al.} \cite{nguyen2016collecting} have pointed that Duchi~\textit{et al.}'s solution doesn't achieve \mbox{$\epsilon$-LDP} when $d$ is even. Nguy{\^e}n~\textit{et al.} have proposed one possible solution to fix Duchi \textit{et al.}'s method to satisfy LDP when $d$ is even. Their method is to re-define a Bernoulli variable $u$ such that
\begin{align}\label{fixing-1}
    \bp{u=1}=\frac{e^\epsilon\cdot C_d}{(e^\epsilon-1)C_d+2^d}.
\end{align}

Furthermore, Nguy{\^e}n~\textit{et al.} proposed Harmony \cite{nguyen2016collecting} that is simpler than Duchi \textit{et al.}'s method when collecting multi-dimensional data with LDP, but achieves the same privacy guarantee and asymptotic error bound. Given an input $V^i$, Harmony returns a perturbed tuple $\widehat{V}^i$ which has non-zero value on only one dimension $j\in[1,d]$. That is, Harmony uniformly at random samples only one dimension $j$ from $[1,d]$ and returns a noisy value $\widehat{v}_j^i$ that is generated from the following distribution
\begin{align}
    \bp{\widehat{v}_j^i = x}=
    \begin{cases}
        \frac{v_j^i\cdot (e^\epsilon -1)+e^\epsilon+1}{2(e^\epsilon+1)},&\text{~if~}x=\frac{e^\epsilon+1}{e^\epsilon-1}\cdot d,\\
        \frac{-v_j^i\cdot (e^\epsilon -1)+e^\epsilon+1}{2(e^\epsilon+1)},&\text{~if~}x=-\frac{e^\epsilon+1}{e^\epsilon-1}\cdot d.
    \end{cases}
\end{align}

As we can see, in Harmony, each user only needs to report one bit to the aggregator. Thus, Harmony has a lower communication overhead of $O(1)$ than Duchi \textit{et al.}'s method, but holds the same error bound of $O\left(\frac{\sqrt{d\log d}}{\epsilon\sqrt{N}}\right)$.

Wang \textit{et al.} \cite{wang2019collecting} further proposed piecewise mechanism (PM) that has lower variance and is easier to implement than Duchi \textit{et al.}'s method. 

We first introduce PM for one dimensional data, that is, $d=1$. Given an input $v^i\in[-1,1]$ of user $i$, the PM outputs a perturbed value $\widehat{v}^i$ in $[-C,C]$, where $C=\frac{e^{(\epsilon/2)}+1}{e^{(\epsilon/2)}-1}$. The probability density function (pdf) of $\widehat{v}^i$ follows a piecewise constant function as
\begin{align}\label{piecewise}
    pdf(\widehat{v}^i=x|v^i)=
    \begin{cases}
        p,&\text{if~}x\in[\ell_{v^i},r_{v^i}],\\
        \frac{p}{e^\epsilon},&\text{if~}x\in[-C,\ell_{v^i}]\cup [r_{v^i},C], 
    \end{cases}
\end{align}
where $p=\frac{e^\epsilon-e^{\epsilon/2}}{2(e^{\epsilon/2+1})}$, $\ell_{v^i}=(C+1)/2\cdot v^i - (C-1)/2$, and $r_{v^i}=\ell_{v^i}+C-1$.

Based on Eq.~(\ref{piecewise}), PM samples a value $u$ uniformly at random from $[0,1]$ and returns $\widehat{v}^i$ uniformly at random from $[\ell_{v^i}, r_{v^i}]$ if $u<\frac{e^{\epsilon/2}}{e^{\epsilon/2}+1}$ (or returns $\widehat{v}^i$ uniformly at random from $[-C,\ell_{v^i}]\cup [r_{v^i},C]$ if $\frac{e^{\epsilon/2}}{e^{\epsilon/2}+1}\leq u\leq 1$).

Wang \textit{et al.} \cite{wang2019collecting} have proved that the variance of PM for one-dimensional data is $\frac{4e^{\epsilon/2}}{3(e^{\epsilon/2}-1)^2}$. Recall that the variance of Duchi \textit{et al.}'s method for one-dimensional data is $\frac{(e^\epsilon+1)^2}{(e^\epsilon-1)^2}$. It can be verified that the variance of PM will smaller than that of Duchi \textit{et al.}'s method when $\epsilon>1.29$.

Furthermore, Wang \textit{et al.} \cite{wang2019collecting} extended the PM for collecting multi-dimensional data based on the idea of Harmony \cite{nguyen2016collecting}. Given an input tuple $V^i\in[-1,1]^d$, it returns a perturbed $\widehat{V}^i$ that has non-zero value on $k$ dimensions at most, where $m=\max\{1,\min\{d,\left \lfloor \frac{\epsilon}{2.5} \right \rfloor\}\}$. In this way, the PM for multi-dimensional data has a error bound of $O\left(\frac{\sqrt{d\log d}}{\epsilon\sqrt{N}}\right)$. Especially, $m$ is much smaller than $d$ and equals to 1 for $\epsilon<5$.

%However, the outputs of PM \cite{wang2019collecting}  and HM \cite{wang2019collecting} may have infinite possibilities, which makes it hard to encode. To solve this problem, Zhao \textit{et al.} \cite{zhao2020local} proposed Three-Outputs mechanism which discretizes the perturbed outputs into three values. Three-Outputs is easy to encode and has low communication cost while providing a smaller worst-cost variance under the high privacy regime (i.e., small privacy budget $\epsilon$). To reduce 

\textit{Comparisons and discussions}. Table~\ref{compare-ldp-mean} summarizes of LDP algorithms for mean estimation on multi-dimensional numeric data. As we can see, Laplace \cite{dwork06Calibrating} and Duchi~\textit{et al.}'s method cost high communication overhead while Harmony \cite{nguyen2016collecting} and PM \cite{wang2019collecting} have low communication costs. For $d$-dimensional data, Laplace has the largest error bound. In contrast, the other three mechanisms have lower error bounds than Laplace. Note that, in theory, PM \cite{wang2019collecting} holds the communication cost of $O(k)$ and error bound of $O\left(\frac{\sqrt{dk\log d}}{\epsilon\sqrt{N}}\right)$, where $k$ is much smaller than $d$ and will be 1 when $\epsilon<5$. 

The last column of Table~\ref{compare-ldp-mean} shows the variance of each mechanism for one-dimensional data. It can be verified that the variance of PM is always smaller than Laplace, but slightly worse than Duchi~\textit{et al.}'s method and Harmony when $\epsilon<1.29$. By observing this, Wang \textit{et al.} \cite{wang2019collecting} proposed to combine PM and Duchi~\textit{et al.}'s method into a new Hybrid Mechanism (HM). They have proved that the worst-case variance of HM for $d=1$ is
\begin{align}\label{var-HM}
    Var_{HM}=
    \begin{cases}
        \frac{e^{\epsilon/2}+3}{3e^{\epsilon/2}(e^{\epsilon/2}-1)}+\frac{(e^\epsilon +1)^2}{e^{\epsilon/2}(e^\epsilon -1)^2},&\text{for~}\epsilon > 0.61,\\
        \frac{(e^\epsilon+1)^2}{(e^\epsilon-1)^2},&\text{for~}\epsilon \leq 0.61.
    \end{cases}
\end{align}
Thus, it can be verified the variance of HM is always smaller than other mechanisms in Table~\ref{compare-ldp-mean}.

\begin{table}[t]
\renewcommand\arraystretch{1.5} %行距
\centering\footnotesize
\caption{Comparisons of mean estimation mechanisms on $d$-dimensional numeric data with LDP}\label{compare-ldp-mean}
\setlength{\tabcolsep}{1.5mm}{
%\fontsize{8pt}{\baselineskip}\selectfont
\begin{threeparttable}
\begin{tabular}{c|c|c|c}
\bottomrule[1pt]
\textbf{Algorithms} &\textbf{Comm. Cost} &\textbf{Error Bound} &\textbf{Variance ($d=1$)}\\ \hline
Laplace \cite{dwork06Calibrating} &$O(d)$ &$O\left(\frac{d\sqrt{\log d}}{\epsilon\sqrt{N}}\right)$  & $\frac{8}{\epsilon^2}$ \\ \hline
Duchi~\textit{et al.} \cite{duchi2018minimax} &$O(d)$  &$O\left(\frac{\sqrt{d\log d}}{\epsilon\sqrt{N}}\right)$  &$\frac{(e^\epsilon+1)^2}{(e^\epsilon-1)^2}$ \\ \hline
Harmony \cite{nguyen2016collecting} &$O(1)$ &$O\left(\frac{\sqrt{d\log d}}{\epsilon\sqrt{N}}\right)$ &$\frac{(e^\epsilon+1)^2}{(e^\epsilon-1)^2}$ \\ \hline
PM \cite{wang2019collecting} &$O(m)$ &$O\left(\frac{\sqrt{d\log d}}{\epsilon\sqrt{N}}\right)$ &$\frac{4e^{\epsilon/2}}{3(e^{\epsilon/2}-1)^2}$ \\ \hline
HM \cite{wang2019collecting} &$O(m)$  &$O\left(\frac{\sqrt{d\log d}}{\epsilon\sqrt{N}}\right)$ &Eq.~(\ref{var-HM}) \\
\toprule[1pt]
\end{tabular}
\end{threeparttable}
}
\end{table}

\subsection{Mean Value Estimation on Evolving Data}\label{sec-mean-envolving}

As pointed in section \ref{sec-fre-envolving}, the privacy leakage will accumulate with the increase of time. This also exists in the mean estimation. Ding \textit{et al.} \cite{ding2017collecting} employed both $\alpha$-point rounding and memoization techniques to estimate mean value of the counter data while ensuring strong privacy protection over time. The basic idea of $\alpha$-point rounding is to discretize the data domain based on a discretization granularity $s$. Ding \textit{et al.} proposed 1-bit mechanism \textsf{1BitMean} for mean estimation. Assume that each user $i$ has a private value $v^i(t) \in[0,r]$ at time $t$. The \textsf{1BitMean} requires that each user reports one bit $b^i(t)$ that is drawn from the distribution
\begin{align}
	b^i(t) = 
	\begin{cases}
	1, \text{~with probability~}\frac{1}{e^\epsilon +1}+\frac{v^i(t)}{r}\cdot\frac{e^\epsilon -1}{e^\epsilon +1},\\
	0, \text{~otherwise}.
	\end{cases}
\end{align}
The mean value of $N$ users at time $t$ can be estimated as
\begin{align}
	\widehat{m}(t) = \frac{r}{N}\sum_{i=1}^{N}\frac{b^i(t)\cdot(e^\epsilon+1)-1}{e^\epsilon-1}.
\end{align}

Based on \textsf{1BitMean}, the procedure of $\alpha$-point rounding includes the following four steps. (\textit{i}) Discretizes data domain $r$ into $s$ parts. (\textit{ii}) Each user $i$ randomly picks a value $\alpha^i \in \{0,1,\cdots,s-1\}$. (\textit{iii}) Each user computes and memoize 1-bit response by invoking \textsf{1BitMean}. (\textit{iv}) Each user performs $\alpha$-rounding based on an arithmetic progression that rounds value to the left is $v^i+\alpha^i <R$, otherwise rounds value to the right. Ding \textit{et al.} \cite{ding2017collecting} have proved that the accuracy of $\alpha$-point rounding mechanism is the same as \textsf{1BitMean} and is independent of the choice of discretization granularity $s$.

\section{Machine Learning with LDP}\label{sec-ml}

Machine learning, as an essential data analysis method, has been applied to various fields. However, the training process may be vulnerable to many attacks (such as membership inference attacks \cite{shokri2017membership}, memorizing model attacks \cite{song2017machine}, model inversion attacks \cite{fredrikson2015model}). For example, adversaries may extract the memorized information in the training process to approximate the sensitive data of the users \cite{song2017machine}. What's worse, Fredrikson \textit{et al.} \cite{fredrikson2015model} have shown an example that the adversary could recover images from a facial recognition system under model inversion attacks, which shows the weakness of a trained machine learning model.

The machine learning algorithms with global DP have been extensively studied by imposing private training \cite{abadi2016deep,phan2017adaptive,lee2018concentrated,zhao2019differential,jayaraman2019evaluating}. With the introduction of LDP, the machine learning algorithms with LDP have been also investigated to achieve privacy protection in a distributed way. The following subsections summarize the existing machine learning algorithms with LDP from the perspective of supervised learning, unsupervised learning, empirical risk minimization, deep learning, reinforcement learning, and federated learning.

\subsection{Supervised Learning}\label{sec-ml-supervised}

Supervised learning algorithms focus on training a prediction model describing data classes via a set of labeled datasets.

Yilmaz \textit{et al.} \cite{yilmaz2019locally} proposed to train a Na\"ive Bayes classifier with LDP. Na\"ive Bayes classification is to find the most probable label when given a new instance. In order to compute the conditional distributions, we need to keep the relationships between the feature values and class labels when perturbing input data. To keep this relationship, Yilmaz \textit{et al.} transformed each user's value and label into a new value first and then performed LDP perturbation. Xue \textit{et al.} \cite{xue2019joint} also aimed at training a Na\"ive Bayes classifier with LDP. They proposed to leverage the joint distributions to compute the conditional distributions. Besides, Berrett and Butucea \cite{berrett2019classification} further considered the binary classification problem with LDP.

High-dimensionality is a big challenge for training a classifier with LDP, which will result in huge time cost and low accuracy. One of the traditional solutions is dimensionality reduction, such as Principal Component Analysis (PCA) \cite{kung2014kernel}. However, the effective dimensionality reduction methods with LDP in machine learning still need further research. Moreover, user partition is always used when learning a model with LDP. For example, the work in \cite{yilmaz2019locally} partitions users into two groups to compute the mean value and squares, respectively. However, simply partitioning users into groups will reduce the estimation accuracy. Therefore, research on supervised learning with LDP still has a long way to go.

\subsection{Unsupervised Learning}\label{sec-ml-cluster}

The problem of clustering has been studied under centralized DP \cite{nissim2016locating,su2016differentially,feldman2017coresets}. With LDP model, Nissim and Stemmer \cite{nissim2018clustering} conducted $1$-clustering by finding a minimum enclosing ball. Moreover, Sun~\textit{et al.}~\cite{sun2019distributed} have investigated the non-interactive clustering under LDP. They extended the Bit Vector mechanism in \cite{karapiperis2017distance,karapiperis2018federal} by modifying the encoding process and proposed kCluster algorithm in an anonymous space based on the improved encoding process. Furthermore, Li~\textit{et al.}~\cite{li2017local} proposed a local-clustering-based collaborative filtering mechanism that utilizes the kNN algorithm to group items and ensure the item-level privacy specification.

For clustering in the local model, the respondent randomizes her/his own data and reports to an untrusted data curator. Although the accuracy of local clustering is not good as that in the central model, local clustering algorithms can achieve stronger privacy protection for users and are more practical for privacy specifications, such as personalized privacy parameters \cite{akter2017computing,nie2019utility}. Xia \textit{et al.} \cite{xia2020distributed} applied LDP on $K$-means clustering by directly perturbing the data of each user. They proposed a budget allocation scheme to reduce the scale of noise to improve accuracy. However, the investigation on clustering under LDP is still in the early stage of research.

\subsection{Empirical Risk Minimization}\label{sec-ml-erm}

In machine learning, the error computed from training data is called empirical risk. Empirical risk minimization (ERM) is such a process that computes an optimal model from a set of parameters by minimizing the expected loss \cite{chaudhuri2011differentially}. A loss function $\mathcal{L}(\theta; x,y)$ is parameterized by $x,y$ and aims to map the parameter vector $\theta$ into a real number. The goal of ERM is to identify a parameter vector $\theta ^*$ such that
\begin{align}
\theta ^* = \mathop{\arg\min}_{\theta}\left[\frac{1}{N}\left(\sum_{i=1}^{N}\mathcal{L}(\theta; x_i,y_i)\right)+\frac{\lambda}{2}\|\theta\|_2^2\right],
\end{align}
where $\lambda>0$ is the regularization parameter.

By choosing different loss functions, ERM can be used to solve certain learning tasks, such as logistic regression, linear regression, and support vector machine (SVM).  The interactive model and non-interactive model under LDP have been discussed in existing literature for natural learning problems \cite{wang2018empirical,wang2019sparse}. Apparently, the interactive model has a better accuracy but leads to high network delay and week privacy guarantee. The non-interactive model is strictly stronger and more practical in most settings. 

Smith~\textit{et al.}~\cite{smith2017interaction} initiated the investigation of interaction in LDP for natural learning problems. They pointed out that for a large class of convex optimization problems with LDP, the server needs to exchange information with each user back and forth in sequence, which will lead to network delays. Thus, they investigated the necessity of the interactivity to optimize convex functions. Smith~\textit{et al.} also provided new algorithms that are either non-interactive or only use a few rounds of interaction. Moreover, Zheng~\textit{et al.}~\cite{zheng2017collect} further proposed more efficient algorithms based on Chebyshev expansion under non-interactive LDP, which achieves quasi-polynomial sample complexity bound.

However, the sample complexity in \cite{smith2017interaction,zheng2017collect} is exponential with the dimensionality and will become not very meaningful in high dimensions. In fact, it is quite common to involve high dimensions in machine learning. Wang~\textit{et al.}~\cite{wang2018empirical} have proposed LDP algorithms with its error bound depending on the Gaussian width, which improves the one in \cite{smith2017interaction}, but the sample complexity still is exponential with the dimensionality. Their following work in \cite{wang2019differentially} improves the sample complexity to be quasi-polynomial. However, the practical performance of these algorithms is still limited. Therefore, for the generalized linear model (GLM), Wang~\textit{et al.}~\cite{wang2019estimating} further proved that when the feature vector of GLM is sub-Gaussian with bounded $\ell_1$-norm, then the LDP algorithm for GLM will achieve a fully polynomial sample complexity. Furthermore, Wang and Xu~ \cite{wang2020principal} addressed the principal component analysis (PCA) problem under the non-interactive LDP and proved the lower bound of the minimax risk in both the low and high dimensional settings.

Moreover, the works in \cite{nguyen2016collecting,wang2019collecting} built the classical machine learning models under LDP in a way of empirical risk minimization (ERM). They solved ERM by stochastic gradient descent (SGD). They consider three common machine learning models: linear regression, logistic regression and SVM classification. The SGD algorithm is adopted to compute the target parameter $\theta$ that holds the minimum (or hopefully) loss. Specifically, at each iteration $t+1$, the parameter vector is updated as $\theta_{t+1} = \theta_{t}-\eta \cdot \nabla\mathcal{L}(\theta_{t};x,y)$, where $\eta$ is the learning rate, $\left \langle x,y \right \rangle$ is a tuple of randomly selected user, $\nabla\mathcal{L}(\theta_{t};x,y)$ is the gradient of loss function $\mathcal{L}(\theta_{t})$ at $\theta_{t}$. With LDP, each $\nabla\mathcal{L}$ is perturbed into a noisy gradient $\nabla\mathcal{L}^*$ by an LDP-compliant algorithm before reporting to the aggregator. That is,
\begin{align}\label{eqn-sgd}
	\theta_{t+1} = \theta_{t}-\eta \cdot \frac{1}{|G|}\sum\nolimits_{i \in G} \nabla\mathcal{L}^*_i,
\end{align}
where $\nabla\mathcal{L}^*_i$ is the perturbed gradient of the user $u_i$ and $|G|$ is the batch size.

When considering the data center network (DCN), Fan~\textit{et al.}~\cite{fan2020privacy} investigated the LDP-based support vector regression (SVR) classification for cloud computing supported data centers. Their method achieves LDP based on the Laplace mechanism. Similarly but differently, Yin~\textit{et al.}~\cite{yin2019local} studied the LDP-based logistic regression classification by involving three specific steps, that is, noise addition, feature selection, and logistic regression. LDP is also applied to online convex optimization to avoid disclosing any parameters while realizing unconstrained adaptive online learning \cite{van2019user}. Besides, Jun and Orabona \cite{jun2019parameter} studied the parameter-free SGD problem under LDP. They proposed BANCO that achieves the convergence rate of the tuned SGD without repeated runs, thus reducing privacy loss and saving the privacy budget.

\subsection{Deep Learning}\label{sec-ml-deep}

Deep learning has played an important role in natural language processing, image classification, and so on. However, the adversaries can easily inject the malicious algorithms in the training process, and then extract and approximate the sensitive information of users \cite{song2017machine,osia2020hybrid}.

Arachchige~\textit{et al.}~\cite{arachchige2019local} proposed an LDP-compliant mechanism LATENT to control privacy leaks in deep learning models (i.e., convolutional neural network (CNN)). Just like other LDP frameworks, LATENT integrates a randomization layer (i.e., LDP layer) to against the untrusted learning servers. A big challenge of applying LDP in deep learning is that the sensitivity is extremely large. In LATENT, the sensitivity is $lr$, where $l$ is the length of the binary string and $r$ is the number of layers of the neural network. To address this, Arachchige~\textit{et al.} further improved OUE \cite{wang2017protocol} where the sensitivity is 2. They proposed the modified OUE (MOUE) that has more flexibility for controlling the randomization of 1s and increasing the probability of keeping 0 bits in their original state. Besides, Arachchige~\textit{et al.} proposed utility enhancing randomization (UER) mechanism that further improves the utility of the randomized binary strings.

What's more, by using the teacher-student paradigm, Zhao~in~\cite{zhao2018distributed} investigated the distributed deep learning model under DP and further considered to allow the personalized choice of privacy parameters for each distributed data entity under LDP. Xu \textit{et al.} \cite{xu2019edgesanitizer} have applied LDP on a deep inference-based edge computing framework to privately build complex deep learning models. Overall, deep learning with LDP is in the early stage of research. Further research is still needed to provide strong privacy, reduce high dimensionality, and improve accuracy.

\subsection{Reinforcement Learning}

Reinforcement learning enables an agent to learn a model interactively, which has been widely adopted in artificial intelligence (AI). However, reinforcement learning is vulnerable to potential attacks, leading to serious privacy leakages \cite{pan2019you}.

To protect user privacy, Gajane \textit{et al.} \cite{gajane2018corrupt} initially studied the multi-armed bandits (MAB) problem with LDP. They proposed a bandit algorithm with LDP aiming at arms with Bernoulli rewards. Afterward, Basu \textit{et al.} \cite{basu2019differential} proposed a unifying set of fundamental privacy definitions for MAB algorithms with the graphical model and LDP model. They have provided both the distribution-dependent and distribution-free regret lower bounds.

As for distributed reinforcement learning, Ono and Takahashi \cite{ono2020locally} proposed a framework Private Gradient Collection (PGC) to privately learn a model based on the noisy gradients. Under the PGC, each local agent reports the perturbed gradients that satisfy LDP to the central aggregator who will update the global parameters. Besides, Ren \textit{et al.} \cite{ren2020multi} mainly investigated on the regret minimization for MAB problems with LDP and proved a tight regret lower bound. They proposed two algorithms that achieve LDP based on Laplace perturbation and Bernoulli response, respectively.

Reinforcement learning plays an important role in AI \cite{arulkumaran2017deep}. LDP is a potential technique to prevent sensitive information from leakage in reinforcement learning. But LDP-based reinforcement learning is still in its infancy.

\subsection{Federated Learning}

Federated learning (FL) is one of the core technologies for the development of a new generation of artificial intelligence (AI) \cite{yang2019federated,yang2019federatedbook,li2020federated}. It provides attractive collaborative learning frameworks for multiple data owners/parties \cite{li2019federated}. Although FL itself can effectively balance the trade-off between utility and privacy for machine learning \cite{zheng2020preserving}, serious privacy issues still occurred when transmitting or exchanging model parameters. Therefore, LDP has been widely adopted in FL systems to provide strong privacy guarantees, such as in smart electric power systems \cite{cao2020ifed} or wireless channels \cite{seif2020wireless}.

The studies in \cite{geyer2017differentially,wei2020federated} have adopted global DP \cite{dwork2014algorithmic} to protect sensitive information in FL. However, since FL itself is a distributed learning framework, LDP is more appropriate to FL systems. Truex \textit{et al.} \cite{truex2020ldp} integrated LDP into a FL system for joint training of deep neural networks. Their method can efficiently handle complex models and against inference attacks while achieving personalized LDP. Besides, Wang \textit{et al.} \cite{wang2020federated} proposed FedLDA that is an LDP-based latent Dirichlet allocation (LDA) model in the setting of FL. FedLDA adopts a novel random response with prior, which ensures that the privacy budget is irrelevant to the dictionary size and the accuracy is greatly improved by an adaptive and non-uniform sampling processing.

To improve the model-fitting and prediction of the schemes, Bhowmick \textit{et al.} \cite{bhowmick2018protection} proposed a relaxed optimal LDP mechanism for private FL. Li \textit{et al.} \cite{li2019differentially} introduced an efficient LDP algorithm for meta-learning, which can be applied to realize the personalized FL. As for federated SGD, LDP has been adopted to prevent privacy leakages from gradients. However, the increase of dimension $d$ will cause the privacy budget to decay rapidly and the noise scale to increase, which leads to poor accuracy of the learned model when $d$ is large. Thus, Liu \textit{et al.} \cite{liu2020fedsel} proposed FedSel that only selects the most important top-$k$ dimensions under the premise of stabilizing the learning process. In addition, Sun \textit{et al.} \cite{sun2020ldp} proposed to mitigate the privacy degradation by splitting and shuffling, which reduces noise variance and improve accuracy.

Recently, Naseri \textit{et al.} \cite{naseri2020toward} proposed an analytical framework that empirically assesses the feasibility and effectiveness of LDP and CDP in protecting FL. They have shown that both LDP and global DP can defend against backdoor attacks, but not do well for property inference attacks.

\section{Applications}\label{sec-application}

This Section concludes the wide applications of LDP for real practice and the Internet of Things.

\subsection{LDP in Real Practice}

LDP has been applied to many real systems due to its powerfulness in privacy protection. There are several large scale deployments in the industry as follows.

\begin{itemize}
	\item \textbf{RAPPOR in Google Chrome.} As the first practical deployment of LDP, RAPPOR \cite{erlingsson2014rappor} is proposed by Google in 2014 and has been integrated into Google Chrome to constantly collect the statistics of Chrome usage (e.g., the homepage and search engine settings) while protecting users' privacy. By analyzing the distribution of these settings, the malicious software tampered with the settings without user consent will be targeted. Furthermore, a follow-up work \cite{fanti2016building} by the Google team has extended RAPPOR to collect more complex statistical tasks when there is no prior knowledge about dictionary knowledge.
	
	%\item \textbf{PROCHLO in Google Chrome.} RAPPOR has processed up to billions of daily data over the last 3 years, but received a limited utility in both theory and practice. Therefore, PROCHLO \cite{bittau2017prochlo} is proposed by the Google team to extend and strengthen the best practices in private data collection and processing. PROCHLO is built on an ESA (Encode, Shuffle, Analyze) architecture and developed based on novel techniques such as Intel's SGX \cite{costan2016intel}, secret sharing \cite{maram2019churp}, and blinding \cite{saarinen2018arithmetic}.
	
	\item \textbf{LDP in iOS and MacOS.} The Apple company has deployed LDP in iOS and MacOS to collect typing statistics (e.g., emoji frequency detection) while providing privacy guarantees for users \cite{apple2017local,thakurta2017emoji,apple2019ios}. The deployed algorithm uses the Fourier transformation and sketching technique to realize a good trade-off between massive learning and high utility.
	
	\item \textbf{LDP in Microsoft Systems.} Microsoft company has also adopted LDP and deployed LDP starting with Windows Insiders in Windows 10 Fall Creators Update \cite{ding2017collecting}. This deployment is used to collect telemetry data statistics (e.g., histogram and mean estimations) across millions of devices over time. Both $\alpha$-point rounding and memoization technique are used to solve the problem that privacy leakage accumulates as time continues to grow. 
	
	\item \textbf{LDP in SAP HANA.} SAP announced that its database management system SAP HANA has integrates LDP to provide privacy-enhanced processing capabilities \cite{kessler2019sap} since May 2018. There are two reasons for choosing LDP. One is to provide privacy guarantees when counting sum average on the database. The other is to ensure a maximum transparency of the privacy-enhancing methods since LDP can avoid the trouble and overhead of maintaining a privacy budget.
\end{itemize}

Many other companies (e.g., Firefox \cite{wang2018locally} and Samsung \cite{nguyen2016collecting}) also plan to build LDP-compliant systems to collect the usage statistics of users while providing strong privacy guarantees. It still needs great efforts to research on large-scale, efficient, and accurate privacy-preserving frameworks while monitoring the behaviors of client devices.

\subsection{LDP in Various Fields}

With the rapid development of the Internet of Things (IoT), multiple IoT applications generate big multimedia data that relate to user health, traffic, city surveillance, locations, etc \cite{lin2017survey}. These data are collected, aggregated, and analyzed to facilitate the IoT infrastructures. However, privacy leakages of users have hindered the development of IoT systems. Therefore, LDP plays an important role in data privacy protection in the Internet of Things.

Usman \textit{et al.} \cite{usman2019paal} proposed a privacy-preserving framework PAAL by combining authentication and aggregation with LDP. PAAL can provide each end-device user with strong privacy guarantee by perturbing the aggregated sensitive information. Ou \textit{et al.} \cite{ou2020singular} also adopted LDP to prevent the adversary from inferring the time-series data classification of household appliances. As a promising branch of IoT, the Internet of Vehicles (IoV) has stimulated the development of vehicular crowdsourcing applications, which also results in unexpected privacy threats on vehicle users. Zhao \textit{et al.} \cite{zhao2020local} adopted both LDP and FL models to avoid sensitive information leakage in IoV applications. Besides, the work in \cite{zhao2019survey} has provided a detailed summary of the applications of LDP in the Internet of connected vehicles.

In what follows, we summarize more specific applications of LDP in the Internet of Things.

\subsubsection{Edge Computing}
LDP itself is a distributed privacy model that can be easily used for providing strict privacy guarantees in edge computing applications. Xu \textit{et al.} \cite{xu2019edgesanitizer} proposed a lightweight edge computing framework based on deep inference while achieving LDP for mobile data analysis. Moreover, Song \textit{et al.} \cite{song2019local} also leveraged LDP models to protect the privacy of multi-attribute data based on edge computing. They have solved the problem of maximizing data utility under privacy budget constrains, which improves the accuracy greatly.

\subsubsection{Hypothesis Testing}
% 假设检验的种类包括：t检验，Z检验，卡方检验，F检验等等。
Many existing studies have looked at the intersection of DP and hypotheses testing \cite{gaboardi2016differentially,cai2017priv,sheffet2017differentially,tong2017gaussian}. The private hypothesis testing under LDP has also been studied in \cite{duchi2018minimax,kairouz2016extremal,sheffet2018locally,gaboardi2018local, gaboardi2019locally,acharya2019test}, including identity and independence testing, Z-test, and distribution testing. %have been studied by several works, albeit restricted to discrete distributions.

Duchi~\textit{et al.}~\cite{duchi2018minimax} initiated to define the canonical hypothesis testing problem and studied the error bound of the probability of error in the hypothesis testing problem. From the perspective of information theory, Kairouz~\textit{et al.}~\cite{kairouz2014extremal,kairouz2016extremal} studied the maximizing of $f$-divergence utility functions under the constraints of LDP and made the effective sample size reduce from $N$ to $\epsilon^2N$ for hypothesis testing. 

Both studies in \cite{sheffet2018locally,gaboardi2018local} investigated hypothesis testing with LDP and presented the asymptotic power and the sample complexity. Sheffet in \cite{sheffet2018locally} showed a characterization for hypothesis testing with LDP and proved the bound of sample complexities for both identity-testing and independence testing when using randomized response techniques. Gaboardi and Rogers \cite{gaboardi2018local} focused on the goodness of fit and independence hypothesis tests under LDP. They designed three different goodness of git tests that utilize different protocols to achieve LDP and guarantee the convergence to a chi-square distribution. Afterward, Gaboardi \textit{et al.} \cite{gaboardi2019locally} further provided upper- and lower-bounds for mean estimation under $(\epsilon,\delta)$-LDP and showed the performance of LDP-compliant Z-test algorithm. Moreover, Acharya~\textit{et al.} in \cite{acharya2019test} have presented the optimal locally private distribution testing algorithm with optimal sample complexity, which improves on the sample complexity bounds in \cite{sheffet2018locally}.

\subsubsection{Social Network}
In the Internet of Things, social network analysis has drawn much attention from many parties. Providing users with effective privacy guarantees is a key prerequisite for analyzing social network data. Therefore, privacy-preserving on social network publishing has been widely investigated with LDP. Usually, a social network is formalized as a graph.

A big challenge is how to apply LDP on the aggregation and generation of the complex graph structures. Qin \textit{et al.} \cite{qin2017generating} initially formulated the problem of generating synthetic decentralized social graphs with LDP. In order to obtain graph information from simple statistics, they proposed LDPGen that incrementally identifies clusters of connected nodes under LDP to capture the structure of the original graph. Zhang \textit{et al.} \cite{zhang2018two} adopted the idea of multi-party computation clustering to generate a graph model under the optimized RR algorithm. Besides, Liu \textit{et al.} \cite{liu2020local} utilized the perturbed local communities to generate a synthetic network that maintains the original structural information.

The studies in \cite{gao2017preserving, gao2018local} focus on online social networks (OSN) publishing with LDP. The key idea of \cite{gao2017preserving, gao2018local} is graph
segmentation, which can reduce the noise scale and the output space. Specifically, they split the original graph to generating the representative hierarchical random graphs (HRG) and then perturbed each local graph with LDP.  Yang \textit{et al.} \cite{yang2020graph} also applied a hierarchical random graph model for publishing hierarchical social networks with LDP. They further employed the Monte Carlo Markov chain to reduce the possible output space and improve the efficiency while extracting HRG with LDP. In addition, Ye \textit{et al.} \cite{ye2020towards} focused on building an LDP-enabled graph metric estimation framework that supports a variety of graph analysis tasks, such as graph clustering, synthetic graph generation, community detection, etc. They proposed to compute the most popular graph metrics only by three parameters, that is adjacency bit vector, adjacency matrix, and node degree.

What's more, social networks are decentralized in nature. In addition to containing her own connections, a participant's local view also contains the connections of her neighbors that are private for the neighbors, but not directly private for the participant herself. In this case, Sun \textit{et al.} \cite{sun2019analyzing} indicated that the general LDP is insufficient to protect the privacy of all participants. Therefore, they formulated a stringent definition of decentralized differential privacy (DDP) that ensures the privacy of both participant and her neighbors. Besides, Wei \textit{et al.} \cite{wei2020asgldp} further proposed AsgLDP for generating decentralized attributed graphs with LDP. Specifically, AsgLDP is composed of two phases that are used for unbiased information aggregation and attributed graph generation, respectively, which preserves the important properties of social graphs.

As for social network analysis, it's hard to obtain a global view of networks since the local view of each user is much limited. The most existing methods solve this challenge by partitioning users into disjoint groups (e.g., \cite{qin2017generating,zhang2018two}). However, the performance of this approach is severely restricted by the number of users. Besides, the computational cost should also be considered for large scale graphs.

\subsubsection{Recommendation System}
The recommendation system makes most of the services and applications in the Internet of Things feasible \cite{jeong2019big}. However, the recommendation system may abuse user data and extract private information when collecting the rating pairs from each user \cite{calandrino2011you}.

While integrating the privacy-preserving techniques into the recommendation system, a big challenge is how to compromise privacy and usability. Liu \textit{et al.} \cite{liu2018when} proposed an unobtrusive recommendation system that balances between privacy and usability by crowdsourcing user privacy settings and generating corresponding recommendations. To provide stronger privacy guarantees, Shin \textit{et al.} \cite{shin2018privacy} proposed LDP-based matrix factorization algorithms that protect both user’s items and ratings. Meanwhile, they utilized the dimensionality reduction technique to reduce domain space, which improves the data utility. Jiang \textit{et al.} \cite{jiang2019towards} proposed a more reliable Secure Distributed Collaborative Filtering (SDCF) framework that ables to preserve the privacy of data items, recommendation model, and the existence of the ratings at the same time. Nonetheless, SDCF performs RAPPOR in each iteration to achieve LDP protection, which will lead to a large perturbation error and low accuracy. To ensure the accuracy, Guo \textit{et al.} \cite{guo2019locally} proposed to reconstruct the collaborative filtering based on similarity scores, which greatly improves the trade-off between privacy and utility.

Although many studies have investigated the recommendation system with LDP, the low accuracy cased by high dimensionality and high sparse rating dataset still remains a huge challenge 
\cite{gao2020dplcf}. Besides, the targeted LDP protocols for recommendation systems need further study.

%\subsubsection{Others}location, genetics, biomedicine,

\section{Discussions and Future Directions}\label{sec-future}

This section presents some discussions and research directions for LDP.

\subsection{Strengthen Theoretical Underpinnings} 

There are several theoretical limitations under LDP needing to be further strengthened. 

\begin{enumerate}
	% [label=(\arabic*)]
	\item The lower bound of the accuracy is not very clear. Duchi \textit{et al.} \cite{duchi2018minimax} have shown a provably optimal estimation procedure under LDP. But the lower bounds on the accuracy of other LDP protocols should be proved elaborately. 
	
	\item Can the sample complexity under LDP be further reduced? One of the limitations of LDP algorithms is that the number of users should be substantially large to ensure the data utility. A general rule of thumb \cite{erlingsson2014rappor} is $\prod_{i=1}^{d}|\Omega_i| \propto \sqrt{N}/10$, where $|\Omega_i|$ is the domain size of the $i$th attribute and $N$ is the data size. Some studies in \cite{acharya2019hadamard,gursoy2019secure} have focused on the scenarios with a small number of users and tried to reduce the sample complexity for all privacy regimes. 
	
	\item There is relatively little research on relaxation of LDP (i.e., $(\epsilon,\delta)$-LDP). It needs to be theoretically studied to show whether we can get any improvements on utility or others when adding a relaxation to LDP \cite{bun2018heavy}. 
	
	\item The more general variant definitions of LDP can be further studied. Some novel and more strict definitions based on LDP have been proposed. But they are only for some specific datasets. For example, $d$-privacy is only for location datasets, and decentralized differential privacy (DDP) is only for graph datasets.
	
\end{enumerate}

\subsection{Overcome the Challenge of Knowing Data Domain.} 

As shown in Section~\ref{sec-fre}, most classical LDP-based frequency estimation mechanisms need to know the domain of attributes in advance, such as PrivSet~\cite{wang2018privset}, LDPMiner~\cite{qin2016heavy}. However, it's unreasonable and hard to make an assumption about the attribute domain. For example, when estimating the statistics of input words of users, the domain of words is very huge and the new word may occur over time. Both studies in \cite{fanti2016building} and \cite{wang2018privtrie} try to address this problem. But the error bounds still depend on the size of character domain and the node number of a trie, which limits the data utility when data domain is large. Thus, how to overcome the challenge of knowing data domain remains challenging.

\subsection{Focus on Data Correlations} 

One of the limitations of existing methods is the neglect of data correlations. Such data correlations appear in multiple attributes \cite{wang2019locallycopula}, repeatedly collected data \cite{erlingsson2014rappor}, or evolving data \cite{ding2017collecting}. These correlations will leak some additional information about users. However, many LDP mechanisms neglect such inadvertent correlations, leading to a degradation of privacy guarantees. Some studies \cite{ye2019privkv,gu2019pckv} have focused on the correlations of key-value data. The researches in \cite{ding2017collecting,joseph2018local} have focused on the temporal correlations of evolving data. However, it still remains open problems of how to learn the data correlations privately and how to integrate such correlations into the encoding principles of LDP protocols.

\subsection{Address High-dimensional Data Analysis} 

The privacy-preserving data analysis on high-dimensional data always suffers from high computation/communication costs and low data utility, which is manifested in two aspects when using LDP. The first is computing joint probability distributions \cite{ren2018LoPub,wang2019locallycopula} (or, $k$-way marginals \cite{cormode2018marginal}). In this case, the total domain size increases exponentially, which leads to huge computing costs and low data utility due to  the``curse of dimensionality''. The second is protecting high-dimensional parameters of learning models in machine learning, deep learning, or federated learning tasks \cite{liu2020fedsel}. In this case, the scale of injected noise is proportional to the dimension, which will inject heavier noise and result in inaccurate models. Therefore, addressing high-dimensional data analysis is an urgent concern in the future.

\subsection{Adopt Personalized/Granular Privacy Constraints} 

Since different users or parties may have distinct privacy requirements, it's more appropriate to design personalized or granular LDP algorithms to protect the data with distinct sensitivity levels. LDP itself is a distributed privacy notion. It can easily achieve personalized/granular privacy protection. Some existing work \cite{wang2015personalized,nie2019utility} aimed to propose personalized LDP-based frameworks for private histogram estimation. Gu \textit{et al.} \cite{gu2020providing} presented Input-Discriminative LDP (ID-LDP) that is a fine-grained privacy notion and reflects the distinct privacy requirements of different inputs. However, adopting personalized/granular privacy constraints still raises further concerns when considering complex system architectures and ensuring good data utility.

\subsection{Develop Prototypical Systems} 

LDP has been widely adopted to deal with many analytic tasks and implemented in many real applications, such as Google Chrome \cite{erlingsson2014rappor}, federated learning systems \cite{truex2020ldp}. However, the prototypical systems based on LDP hardly appeared at present. By developing prototypical systems, we can further improve the LDP algorithms based on user requirements and the running results in an interactive way.

\section{Conclusion}\label{sec-conclusion}

Data statistics and analysis have greatly facilitated the progress and development of the information society in the Internet of Things. As a strong privacy model, LDP has been widely adopted to protect users against information leaks while collecting and analyzing users' data. This paper presents a comprehensive review of LDP, including privacy models, data statistics and analysis tasks, enabling mechanisms, and applications. We systematically categorize the data statistics and analysis tasks into three aspects: frequency estimation, mean estimation, and machine learning. For each category, we summarize and compare the state-of-the-art LDP-based mechanisms from different perspectives. Meanwhile, several applications in real systems and the Internet of Things are presented to demonstrate how LDP to be implemented in real-world scenarios. At last, we explore and conclude some future research directions from several perspectives.

\ifCLASSOPTIONcaptionsoff
\newpage
\fi

\bibliographystyle{IEEEtran}
\bibliography{IEEEabrv,related}

\end{document}